\documentclass[a4paper]{JHEP3}
\usepackage{epsfig, subfigure}
\usepackage{graphicx}
\usepackage{amsmath}

\usepackage{mcite}

\newcommand{\ms}{\overline{MS}}
\newcommand{\nn}{\nonumber}

\renewcommand{\Re}{\mbox{Re}}
\providecommand{\Li}[1]{\text{Li}\,_2\left(#1\right)}

\newcommand{\qqbar}{q\overline{q}}

\newcommand{\Oa}{{\cal O}(\alpha)}
\newcommand{\Oas}{{\cal O}(\alpha^{\mathrm 2})}
\newcommand{\Oaass}{{\cal O}(\alpha\alpha_{\mathrm s}^{\mathrm 2})}
\newcommand{\Oaas}{{\cal O}(\alpha\alpha_{\mathrm s})}
\newcommand{\Oass}{{\cal O}(\alpha_{\mathrm s}^{\mathrm 2})}
\newcommand{\Oasc}{{\cal O}(\alpha_{\mathrm s}^{\mathrm 3})}
\newcommand{\Oasf}{{\cal O}(\alpha_{\mathrm s}^{\mathrm 4})}

\def \pt{p_T}
\def \tb{\tan\beta}
\def \cm{c.\,m.\,}
\def \msu{m_{\tilde{U}_3}}
\def \msq{m_{\tilde{Q}_3}}
\def \qq{q\overline{q}}
\def \gy{g\gamma}
\def \mstop{m_{\tilde{t}_1}}
\def \mstopi{m_{\tilde{t}_i}}
\def \lnbeta{\ln \left(\frac{1+\beta}{1-\beta}\right)}

\renewcommand{\eqref}[1]{Eq.~(\ref{#1})}

\newcommand{\figref}[1]{Fig.~\ref{#1}}

\providecommand{\pict}[3]{\scalebox{#1}{\rotatebox{#2}{\epsffile{#3}}}}

\newcommand{\tr}{\,\tilde{t}_R}
\newcommand{\tl}{\,\tilde{t}_L}
\newcommand{\stst}{\,\tilde{t}\tilde{t}^*}
\newcommand{\ststi}{\,\tilde{t}_i\tilde{t}_i^*}

\newcommand{\ti}{\,\tilde{t}_{i}}

\newcommand{\tone}{\,\tilde{t}_1}
\newcommand{\ttwo}{\,\tilde{t}_2}

\providecommand{\neut}[1]{\tilde{\chi}_{#1}^0}

\makeatletter
\renewcommand\appendix{\par
  \setcounter{section}{0}%
  \setcounter{subsection}{0}%
  \renewcommand\thesection{\@Alph\c@section}
  \setcounter{figure}{0}%
  \setcounter{table}{0}%
  \renewcommand\thefigure{\thesection.\@arabic\c@figure}
  \renewcommand\thetable{\thesection.\@arabic\c@table}
  }
\makeatother

\title{Hadronic production of top-squark pairs with electroweak NLO contributions}

\author{Wolfgang Hollik, Monika Koll\'ar, and  Maike K. Trenkel \\
Max-Planck-Institut f\"ur Physik, F\"ohringer Ring 6, D-80805 M\"unchen, Germany}

\abstract{Presented are complete next-to-leading order electroweak (NLO EW) 
corrections to top-squark pair production at the Large Hadron Collider (LHC) 
within the Minimal Supersymmetric Standard Model (MSSM). At this order, also effects 
from the interference of EW and QCD contributions have to be taken into account. 
Moreover, photon-induced top-squark production is considered as an additional 
partonic channel, which arises from the non-zero photon density in the proton.

\bigskip

}

\preprint{MPP-2007-149 \\ arXiv:0712.0287 [hep-ph]}

\begin{document}

\section{Introduction}

Within supersymmetric theories top-squarks are the supersymmetric partners of
the left- and right-handed top quarks. 
The two superpartners $\tl$ and $\tr$, which belong
to chiral supermultiplets $\hat{Q}$ and $\hat{T}$, 
in general mix to produce two mass 
eigenstates $\tone$ and $\ttwo$. 
In many supersymmetric models the lighter mass eigenstate appears as
the lightest colored particle~\cite{Ellis:1983ed}, for reasons related to
the large top Yukawa coupling.
The large mixing in the stop sector leads to a substantial
splitting between the two mass eigenstates, 
and the evolution from the GUT scale
to the electroweak  scale yields low values for the stop masses 
when a universal
scalar mass is assumed at the high  scale~\cite{Djouadi:1996pj}.
The search for top-squarks is therefore of particular interest 
for the coming LHC experiments, 
where they would be primarily produced in pairs via the strong interaction, 
with relatively large cross sections.

Current experimental limits on top-squark pair production include 
searches performed 
at LEP \cite{Heister:2002hp,*Abreu:2000qj,*Acciarri:1999xb,*Abbiendi:2002mp} 
reviewed e.g~in~\cite{Kraan:2003it}, and at the Tevatron, done by the CDF and
D\O~collaborations in approximately 90~pb$^{-1}$ of Run~I data 
\cite{Affolder:1999cz,*Abazov:2004ca}. 
Extended searches have been done using Run~II data samples by both 
CDF and D\O~\cite{Aaltonen:2007sw,*Abazov:2006wb,*Abazov:2006bj}.
Limits on the top-squark mass, depending on the mass of the lightest neutralino,
are provided with the assumption that $BR( \tone \rightarrow c \neut{1} )=100$~\% in
\cite{Nunnemann:2006sf}.

Experimental searches for the top-squarks have also been done in $ep$ collisions at 
HERA~\cite{Aktas:2004tm,*unknown:2006je}, where only single stop production 
could be kinematically accessed and hence 
constraints have been derived essentially on the 
R-parity violating class of supersymmetric models.

Concerning the theoretical predictions, 
QCD-based Born-level cross sections for the production 
of squarks and gluinos in hadron collisions have been calculated 
in \cite{Kane:1982hw,*Harrison:1982yi,*Reya:1984yz,*Dawson:1983fw,*Baer:1985xz}. 
They have been improved by including NLO corrections 
in supersymmetric QCD (SUSY-QCD), worked out in
\cite{Beenakker:1994an,*Beenakker:1996ch} with the restriction to final state 
squarks of the first two generations,
and for the stop sector in~\cite{Beenakker:1997ut}. 
The production of top-squark pairs in hadronic collisions is diagonal at 
lowest order at $\Oass$.
Electroweak (EW) contributions of $\Oas$  
 are suppressed by two orders of magnitude.
Also at $\Oasc$ the production mechanism is still diagonal.
Non-diagonal production occurs at $\Oasf$, and the cross section
is accordingly suppressed.
Production of non-diagonal top-squark
pairs can also proceed at $\Oas$ 
mediated by $Z$-exchange
through $\qqbar$ annihilation~\cite{Bozzi:2005sy}    
as well as in $e^+e^-$ annihilation~\cite{Bartl:1997yi}.

The LO cross section for diagonal top-squark pair production depends only 
on the mass of the produced squarks. As a consequence, bounds on the 
production cross section can easily be translated into lower 
bounds on the lightest stop mass. At NLO, the cross section becomes considerably 
changed and dependent on other supersymmetric parameters, like mixing angles, 
gluino mass, masses of other squarks, etc., which enter through the higher 
order terms. Once top-squarks are discovered, 
measurement of their masses and cross sections will provide 
important observables for testing and constraining the supersymmetric model. 

In the following, we study the NLO contributions to 
diagonal top-squark pair production
that arise from the electroweak interaction
within the Minimal Supersymmetric Standard Model (MSSM). 
We assume the
MSSM with real parameters,  R-parity conservation, and minimal flavor violation.
The outline of our paper is as follows. In Section~2, we present
analytical expressions for the partonic and hadronic LO cross sections. 
We also introduce some basic notations used throughout the paper. Section~3
is dedicated to the classification of the NLO EW contributions into virtual and
real corrections with the treatment of soft and collinear singularities,
and photon-induced contributions.
In Section~4, we give a list of input parameters 
and conventions, followed by our numerical results for the hadronic 
cross sections and distributions for $pp$ collisions at a center-of-mass energy
$\sqrt{S}=14$~TeV at the LHC. We also investigate the application of kinematical
cuts, and we analyze the impact of varying the MSSM parameters.

\section{Top-squark eigenstates and LO cross sections}

In the MSSM Lagrangian, mixing of the left- and right-handed top-squark eigenstates 
$\tilde{t}_{L/R}$ into mass eigenstates $\tilde{t}_{1/2}$ is induced by the  trilinear 
Higgs-stop-stop coupling term $A_t$ and the Higgs-mixing parameter $\mu$. 
The top-squark mass matrix squared is given by \cite{Haber:1984rc}
\begin{align}
         \mathfrak{M}^2 &= \left( {m_t^2 + A_{LL} \qquad m_t B_{LR}} \atop {m_t B_{LR}
\qquad m_t^2 + C_{RR} } \right),
\end{align}
with $m_t$ denoting the top-quark mass and 
\begin{align}
\begin{split}
        A_{LL}  &=  \Big(\frac{1}{2} - \frac{2}{3} \sin^2{\theta _W} \Big)\, m_Z^2\,  \cos{2 \beta} + \msq^2\,, \\
        B_{LR}  &= A_t - \mu \cot{\beta}\,, \\
        C_{RR} &= \frac{2}{3} \, \sin^2{\theta_W} \, m_Z^2\, \cos{2 \beta} + \msu^2\,.
\end{split}
\end{align}
Here, $\tb$ is the ratio of the vacuum expectation values of the two Higgs doublets and 
$\msq,\, \msu$ are the soft-breaking mass terms for 
left- and right-handed top-squarks, respectively. 

The top-squark mass eigenvalues are obtained by diagonalizing the mass matrix,
\begin{align} 
        &U \mathfrak{M}^2 U^{\dagger} \, = \, \left({ m_{\tilde{t}_1}^2 \quad 0} \atop {0 \quad m_{\tilde{t}_2}^2}
                \right) , 
        \qquad U = \left({{\, \cos \theta_{\tilde t} \quad \, \sin \theta_{\tilde t}} 
                \atop {-\sin \theta_{\tilde t} \quad \cos \theta_{\tilde t} }} \right),
\\
        m_{\tilde{t}_{1,2}}^2 & =
                m_t^2 +\,  \frac{1}{2} \, \bigg( A_{LL} + C_{RR} \mp \sqrt{(A_{LL}-C_{RR})^2
                + 4 m_t^2 B_{LR}^2}\, \bigg),
\label{eq_stopmass}
\end{align}
and the mixing angle $\theta_{\tilde t}$ is determined by
\begin{align}
        \tan{2 \theta_{\tilde t}} &=  \frac{2 m_t B_{LR}}{A_{LL} -C_{RR}} .
\end{align}

At hadron colliders, diagonal pairs of top-squarks can be produced 
at leading order in QCD in two classes of partonic subprocesses,
\begin{align}
\begin{split}
        gg \rightarrow \tone\tone^* & \quad \mbox{and} \quad  \ttwo\ttwo^*, \\ 
        q\bar{q} \rightarrow \tone\tone^* & \quad \mbox{and} \quad  \ttwo\ttwo^*\,,
\label{eq_processes}
\end{split}
\end{align}
where $\qqbar$ denotes representatively the contributing quark flavors.
The corresponding Feynman diagrams for the example of $\tone\tone^*$ production 
are shown in the appendix, \figref{fig_born-beide}.
As already mentioned, mixed pairs cannot be produced at lowest order since the
$g\stst$ and $gg\stst$ vertices are diagonal in the chiral as well as in the 
mass basis. 

The differential partonic cross sections for the subprocesses, 
\begin{align}
        d\hat \sigma_0^{gg, q\bar{q}} (\hat s) =
        \frac{1}{16 \pi \hat s^2}\,\overline{\sum} 
        \bigl\lvert{\cal M}^{gg, q\bar{q}}_0 (\hat s,\hat t,\hat u) \bigr\rvert^2 
        d\hat{t}\; , 
\end{align}
can be expressed in terms of the squared and spin-averaged lowest-order
matrix elements, as explicitly given by~\cite{Beenakker:1994an,*Beenakker:1996ch},
\begin{align}
        \overline{\sum} \bigl\lvert{\cal M}_0^{gg}\bigr\rvert^2 & = 
                \frac{1}{4} \cdot \frac{1}{64}\cdot 32 \pi^2 \alpha_s^2 
                \left[ C_0\left( 1-2\frac{\hat{t}_r \,\hat{u}_r}{\hat s^2}\right)-C_K  \right]
                \left[ 1-2\frac{\hat s \mstopi^2}{\hat{t}_r \,\hat{u}_r} 
                        \left( 1-\frac{\hat s \mstopi^2}{\hat{t}_r \,\hat{u}_r} \right)\right] ,
\\
        \overline{\sum} \bigl\lvert{\cal M}_0^{q\bar{q}}\bigr\rvert^2 & = 
                \frac{1}{4} \cdot \frac{1}{9}\cdot 64 \pi^2 \alpha_s^2 \, N C_F \;
                \frac{\hat{t}_r \,\hat{u}_r -\mstopi^2\hat s}{\hat s^2} ,
\end{align}
with $\hat{t}_r = \hat t - \mstopi^2,\, \hat{u}_r = \hat u -\mstopi^2$, where
$\hat s,\, \hat t,\, \hat u$ are the usual Mandelstam variables. $i= 1,\,2$ 
denotes the two mass eigenstates.
The $SU(3)$ color factors are given by $N=3$, $C_0 = N(N^2-1)=24$, 
$C_K = (N^2-1)/N=8/3$ and $C_F = (N^2-1)/(2N)=4/3$.

The differential cross section at the hadronic level for the process 
$AB \rightarrow \ststi$, $i=$1, 2, is related to the partonic cross sections through
\begin{align}
        d\sigma^{AB}(S) = \sum_{a,b} \,
                 \int_{\tau_0}^1  \! d\tau \,
                \, \frac{d\mathcal{L}_{ab}^{AB}}{d\tau} 
                \, d\hat{\sigma}_0^{ab} (\hat s)\,, 
\label{eq_formula_hadCS}
\end{align}
with $\tau = \hat{s}/S$, $S\;(\hat{s})$ being the hadronic (partonic) 
center-of-mass energy squared and $\tau_0 = 4\mstopi^2/S$ is the production threshold.
The sum over $a,b$ runs over all possible initial partons.
The parton luminosities are given by
\begin{align}
        \frac{d\mathcal{L}^{AB}_{ab}}{d\tau} = 
                \frac{1}{1+\delta_{ab}} \,      \int_{\tau}^1 \! \frac{dx}{x} 
                \, \biggl[      f_{a/A}\bigl(x, \mu_F\bigr) \,
                                f_{b/B}\Bigl(\frac{\tau}{x}, \mu_F\Bigr)\,
                +               f_{b/A}\Bigl(\frac{\tau}{x}, \mu_F\Bigr) \,
                                f_{a/B}\bigl(x, \mu_F\bigr) \biggr],
\label{eq_luminosity}
\end{align}
where the parton distribution functions (PDFs) $f_{a/A}(x,\mu_F)$ 
parameterize the probability of finding a parton $a$ inside a hadron $A$ 
with fraction $x$ of the hadron momentum at a factorization scale $\mu_F$.

\section{Classification of EW NLO corrections}

In the following we describe the calculation of EW contributions 
to top-squark pair production at NLO.
For the treatment of the Feynman diagrams and corresponding amplitudes 
we make use of
{\tt FeynArts~3.2}~\cite{Kublbeck:1990xc,*Hahn:2000kx,*Hahn:2001rv} and
{\tt FormCalc~5.2}
with {\tt LoopTools~2.2}~\cite{Hahn:1998yk,*Hahn:2006qw}, 
based on Passarino-Veltman reduction techniques for the tensor loop 
integrals \cite{tHooft:1978xw,*Passarino:1978jh},
which were further developed for 4-point  integrals 
in~\cite{Beenakker:1988jr,*Denner:1991qq}.
Higgs properties are computed with 
{\tt FeynHiggs~2.5.1}~\cite{Heinemeyer:1998yj,*Hahn:2006np}.

The supersymmetric final state does not allow to separate the SM-like 
corrections from the superpartner contributions which are necessary for
the cancellation of ultra-violet (UV) singularities. 
As the photino is not a mass eigenstate of the theory, it is also not possible
to split the EW corrections into a QED and a weak part, which is often the case in 
SM processes.
In order to obtain a UV finite result, we have to deal with the complete set
of EW virtual corrections including photonic contributions.
These are infrared (IR) singular and thus also the real photonic corrections have to be 
taken into account. In addition, a photon-induced subclass of corrections 
appears at NLO as an independent production channel.

\subsection{Virtual corrections}

The virtual corrections arise from self-energy, vertex, box, and counter-term
diagrams. These are shown in the appendix,
in~\figref{fig_uu-virtualcorrs} for the $\qqbar$ annihilation and 
and in~\figref{fig_gg-virtualcorrs} for the gluon fusion channel, respectively. 
Getting an UV finite result requires renormalization of the involved quarks and
top-squarks. The renormalized quark and squark self-energies are obtained
from the unrenormalized initial quark self-energies
\begin{equation}
\Sigma^q (p\!\!\!/) = p\!\!\!/\omega_-\Sigma^q_L(p^2) + 
p\!\!\!/\omega_+\Sigma^q_R(p^2) + m_q\Sigma^q_S(p^2),
\end{equation}
according to
\begin{eqnarray}    
\hat \Sigma^q_L (p^2) & = & \Sigma^q_L (p^2) + \delta Z^q_L\,, \nonumber \\  
\hat \Sigma^q_R (p^2) & = & \Sigma^q_R (p^2) + \delta Z^q_R\,, \\  
\hat \Sigma^q_S (p^2) & = & \Sigma^q_S (p^2) - \frac{1}{2} \left( \delta Z^q_L  
+ \delta Z^q_R  \right) + \frac{\delta m_q}{m_q}\,,  \nn
\end{eqnarray}
and from  the top-squark self-energies
$\Sigma_{\ti} (k^2)$ (for $i=$1, 2), 
according to
\begin{equation}  
\hat{\Sigma}_{\ti} (k^2) = \Sigma_{\ti} (k^2) + k^2  
\delta Z_{\ti} - m^2_{\ti}  \delta Z_{\ti} - \delta m^2_{\ti} \; ,
\end{equation}
with the renormalized quantities denoted by the symbol $\hat\Sigma$.

The full set of virtual contributions is UV finite after including 
the proper
counter-terms for self-energies, quark vertices, and squark triple and 
quartic vertices, as listed in the following set of Feynman rules:
\vspace{5mm} \newline
\begin{minipage}[l]{0.3\linewidth}%
\hspace{.18\linewidth}\includegraphics[height=3.5cm]{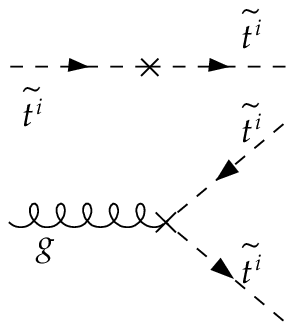}%
\end{minipage}
\begin{minipage}[r]{0.69\linewidth}\vspace{-11mm}
\begin{eqnarray}   
i\delta\Sigma_{\ti} & = & i \left( k^2\,\delta   
Z_{\ti} - m^2_{\ti} \delta Z_{\ti} -\delta   
m^2_{\ti} \right) \; , \\[1cm]  
i\delta\Lambda_{\mu_{i}} & = & -i g_s T^c \left(k + k' \right)_\mu  
\,\delta Z_{\ti} \; ,   
\label{eq:stop_counter1}   
\end{eqnarray}
\end{minipage}
\vspace{4mm}
\newline
\begin{minipage}[l]{0.3\linewidth}%
\hspace{.18\linewidth}\includegraphics[height=4.2cm]{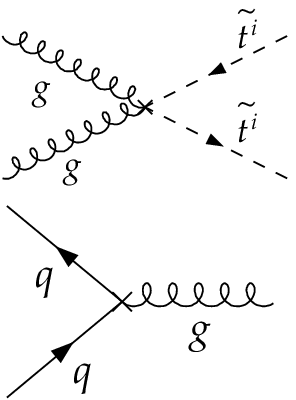}%
\end{minipage}
\begin{minipage}[r]{0.69\linewidth}\vspace{-8mm}
\begin{eqnarray}   
i\delta\Lambda^{SSVV}_{\mu_{i}} & = & \frac{1}{2}i g^2_s \left( \frac{1}{3}  
\delta_{ab} + d_{abc} T^c \right)g_{\mu\nu}  
\,\delta Z_{\ti} \; , \\[1.2cm]  
i\delta\Lambda^q_\mu & = & -i g_s T^c \gamma_\mu  
\,\left( \omega_- \delta Z^q_L + \omega_+ \delta Z^q_R \right) \; ,   
\label{eq:stop_counter2}   
\end{eqnarray}
\end{minipage}
\vspace{4mm}
\newline
where $k$, $k'$ denote the momenta of top-squarks (in the direction of
arrows), $a$, $b$, and $c$ are the
gluonic color indices, $T^c$ and $d_{abc}$ are the color factors (we skip
the fermionic and sfermionic color indices), 
and $\omega_{\pm} = (1\pm \gamma_5)/2$ are the projection operators.
The renormalization constants are fixed within the on-shell renormalization
scheme as follows,
\begin{eqnarray}
\delta m^2_{\ti}  = & & \Re \, \Sigma_{\ti} (m^2_{\ti}) \, ,
\\
\delta Z_{\ti} = &  - & \frac{d}{d k^2} \Re \,
\Sigma_{\ti} (k^2) \bigg|_{k^2=m^2_{\ti}} \, ,
\\
\delta Z^q_{L,R} = & - & \Re \,\Sigma^q_{L,R}(m^2_q) - m^2_q 
\frac{\partial}{\partial p^2} \Re \left[ \Sigma^q_L(p^2) +
\Sigma^q_R(p^2) + 2 \Sigma^q_S(p^2) \right] \bigg|_{p^2=m^2_q} \, .
\label{eq:rcs}
\end{eqnarray}

There is no renormalization of the gluon field at $\Oa$. 
Also, the strong coupling constant does not need renormalization
since UV singularities cancel
in the sum of 3- and 4-point functions and their corresponding counter-terms
from quark and squark field renormalization 
(see Figs.~\ref{fig_uu-virtualcorrs} and \ref{fig_gg-virtualcorrs}
in the appendix).

Loop diagrams involving virtual photons generate IR singularities.
According to Bloch-Nordsieck \cite{Bloch:1937pw}, IR singular terms cancel 
against their counterparts in the real photon corrections. To regularize 
the IR singularities we introduce a fictitious photon mass $\lambda$.
In case of external light quarks, also collinear singularities occur if
a photon is radiated off a massless quark in the collinear limit.
We therefore keep non-zero initial-state quark masses $m_q$ in the loop
integrals. This gives rise to single and double logarithmic contributions 
of quark masses. The double logarithms cancel in the sum of virtual and real 
corrections, single logarithms, however, survive and have to be treated by 
means of the factorization.

In the $gg$ fusion channel, IR singularities originate only
from final-state photon radiation, and mass singularities do not occur.
In the $\qqbar$ annihilation subprocess, the IR singular structure is extended
by the contributions related to the gluons which appear in the 4-point UV
finite loop integrals. There are two types of IR singular box contributions
(\figref{fig_uu-virtualcorrs} c). The first group is formed by the gluon--photon
box diagrams with two sources of IR~singularities, one related to photons,
the other to gluons. The second group consists of the gluon--$Z$ box
diagrams with IR singularities originating from the gluons only.
There is also an IR finite group of $\Oaas$ box diagrams which consists of
gluino--neutralino loops (\figref{fig_uu-virtualcorrs}~d).
Owing to the photon-like appearance of the gluon in the
box contributions, the gluonic IR singularities can be handled in 
analogy to the photon IR singularities.

\subsection{Real corrections}

To compensate IR singularities in the virtual EW corrections, 
contributions with real photon 
(\figref{fig_photonrad}~a and~c) 
and real gluon radiation are required. In case of  $gg$ fusion, only 
photon bremsstrahlung is needed, whereas in the $\qqbar$ annihilation channel,
also gluon bremsstrahlung at the appropriate order $\Oaass$ has to be 
taken into account (\figref{fig_gluonrad}) to cancel the IR singularities related 
to the gluon.
The necessary contributions originate from the interference of QCD and EW Born 
level diagrams, which vanishes at LO.
Not all of the interference terms contribute. Due to the color structure, only
the interference between initial and final state gluon radiation is non-zero.

Including the EW--QCD interference in the real corrections does not yet lead to an 
IR finite result. Also the IR singular QCD-mediated box corrections interfering
with the $\Oa$ photon and $Z$-boson  tree-level diagrams are needed. 
Besides the gluonic corrections there are also the 
IR finite QCD-mediated box corrections, which contain gluinos in the loop.
Interfered with the $\Oa$ tree-level diagrams, these also give contributions
of the respective order of $\Oaass$.
The set of all  $\Oass$ diagrams is shown in \figref{fig_uu-qcdBox}.

So far we have mentioned only the IR singular bremsstrahlung contributions. 
However, there are also IR finite real corrections to both gluon fusion and 
$\qqbar$ annihilation processes. In addition to the photon radiation off the 
off-shell top-squark
there are photon radiation contributions originating from the quartic
gluon--photon--squark--squark coupling. These contributions do not have to 
be regularized since they are not singular (\figref{fig_photonrad} b and d).

The treatment of IR singular bremsstrahlung is done using the
phase space slicing method. 
Imposing cut-offs $\Delta E$ on the photon/gluon energy 
and $\Delta \theta$ on the angle between the photon/gluon and radiating fermion,
the photonic/gluonic phase space
is split into soft and collinear parts which contain singularities
and a non-collinear, hard part which is free of singularities
and is integrated numerically.
 The sum of virtual and real contributions,
each of them dependent on the cut-off parameters
$\Delta E$ and $\Delta \theta$, has to provide a fully 
independent result. To ensure this we perform numerical checks.

In the singular regions, the squared matrix elements for the radiative 
processes factorize into the lowest-order squared matrix elements and 
universal factors containing the singularities.

\subsubsection{Soft singularities}

The soft-photon part of the radiative cross section
in the $\qqbar$ annihilation channel
\begin{align}
        d\hat{\sigma}_{soft,\gamma}^{q\bar{q}} (\hat s) & = \frac{\alpha}{\pi}\,\Big(
                 e_q^2\,\delta_{soft}^{in} + e_t^2\,\delta_{soft}^{fin} 
                + 2 e_q e_t \, \delta_{soft}^{int} \Big) 
                        \,d\hat{\sigma}_0^{q\bar{q}} (\hat s)\,,
\label{eq_sigma_soft_phot}
\end{align}
and in  the $gg$ fusion channel
\begin{align}
        d\hat{\sigma}_{soft,\gamma}^{gg} (\hat s) & = \frac{\alpha}{\pi}\,
                  e_t^2\,\delta_{soft}^{fin} \,d\hat{\sigma}_0^{gg} (\hat s)\,,
\end{align}
can be expressed using universal factors, $\delta_{soft}^{in, fin, int}$,
which refer to the initial state radiation, final state radiation or
interference of initial and final state radiation, respectively. 
$d\hat{\sigma}_0^{q\bar{q} ,gg}$
denote the corresponding partonic lowest order cross sections.
The singular universal factors, similar to those in~\cite{Beenakker:1991ca}, 
read  as follows, 
\begin{align}
\begin{split}
        \delta_{soft}^{in} = & \left[ \ln \delta_s^2 - \ln\frac{\lambda^2}{\hat{s}} \right]
                \left[ \ln \frac{\hat{s}}{m_q^2} - 1 \right] 
                - \frac{1}{2}\, \ln^2 \frac{\hat{s}}{m_q^2}
                +  \ln\frac{\hat{s}}{m_q^2} - \frac{\pi^2}{3}\,,
\\[1.5ex]
        \delta_{soft}^{fin} = &  \left[ \ln \delta_s^2 - \ln\frac{\lambda^2}{\hat{s}} \right]
                \bigg[ \frac{\hat{s}-2\mstopi^2}{\hat{s} \beta} \, \lnbeta - 1 \bigg] 
                + \frac{1}{\beta}\, \lnbeta
 \\
                & - \frac{\hat{s} -2\mstopi^2}{\hat{s} \beta}\,\left[
                         2 \Li{\frac{2\beta}{1+\beta}} + \frac{1}{2} \ln^2
                                 \left(\frac{1+\beta}{1-\beta}\right) \right],
\\[1.5ex]
        \delta_{soft}^{int} = &  \left[ \ln \delta_s^2 - \ln\frac{\lambda^2}{\hat{s}} \right]
                \ln\left( \frac{1-\beta\cos\theta}{1+\beta\cos\theta}\right) 
                - \Li{1-\frac{1-\beta}{1-\beta\cos\theta}}
\\
                & - \Li{1-\frac{1+\beta}{1-\beta\cos\theta}}
                 + \Li{1-\frac{1-\beta}{1+\beta\cos\theta}}
                + \Li{1-\frac{1+\beta}{1+\beta\cos\theta}} .
\label{eq_delta_soft}
\end{split}
\end{align}
Here, $e_q$ and $e_t$ are the electric charges of the initial quark and 
of the top-squark, respectively, 
and we introduced $\delta_s = 2\Delta E / \sqrt{\hat{s}}$, 
where $\Delta E$ is the slicing parameter for the maximum energy a soft 
photon may have. For application purposes, it is useful to express 
\eqref{eq_delta_soft} in terms of Mandelstam invariants, $\hat{t}$ and 
$\hat{u}$, using the relations
\begin{align}
        \hat{t},\, \hat{u} &= \mstopi^2 - \frac{\hat{s}}{2}\, 
                \left( 1 \mp \beta \cos\theta \right),
\qquad
        \beta = \sqrt{ 1 - \frac{4 \mstopi}{\hat{s}}}.
\end{align}

The soft-gluon part for the $\qqbar$ channel can be written in a way
similar to (\ref{eq_sigma_soft_phot}),
but with a different arrangement of the color matrices,
\begin{align} 
\begin{split}
   d\hat{\sigma}_{soft,g}^{q\bar{q}} (\hat s)  =\, &
        \frac{\alpha_s}{\pi}\, \delta_{soft}^{int} \, 
      \Big[T^a_{ij} T^b_{ji}T^a_{lm} T^b_{ml}\Big]\, 
\\       & \times
 2 \,\Re \,\mathrm{\overline{\sum}} 
       \left(\widetilde{\mathcal{M}}_{0,g}^{q\bar{q} \, *} 
             \widetilde{\mathcal{M}}_{0,\gamma}^{q\bar{q}}
+ \widetilde{\mathcal{M}}_{0,g}^{q\bar{q} \, *} 
             \widetilde{\mathcal{M}}_{0,Z}^{q\bar{q}}\right)\, 
      \frac{d\hat{t}}{16\pi \hat{s}^2} \, ,
\label{eq_sigma_soft_gluon} 
\end{split}
\end{align}
with $\widetilde{\mathcal{M}}$ denoting the ``Born'' matrix elements
for $g$, $\gamma$ and $Z$ exchange where the color matrices
are factorized off. Explicitly, it can be written as follows,
\begin{align}
\begin{split}
           d\hat{\sigma}_{soft,g}^{q\bar{q}} (\hat s)  =\, &
        \frac{\alpha_s}{\pi}\, \delta_{soft}^{int} \, N C_F \, 
          \Bigg[ \frac{8 e_q e_t}{\hat{s}^2} + 
            \frac{\big( (U_{1i})^2 -2 e_t \sin^2\theta_W\big)
                  \big( \epsilon -4e_q \sin^2\theta_W\big)}
                {\sin^2\theta_W \cos^2\theta_W\,\hat{s}(\hat{s}-m_Z^2)}\Bigg]\, 
\\      & \times
      \frac{16 \pi^2 \alpha \alpha_s}{4\cdot 9}\,
         \Big[ \big(\hat{t}-\mstopi^2\big) \big(\hat{u}-\mstopi^2\big)
                 - \mstopi^2 \hat{s}\Big]\, 
      \frac{d\hat{t}}{16\pi \hat{s}^2} \,,
\end{split}
\end{align}
involving the top-squark mixing matrix of \eqref{eq_stopmass},
and $\epsilon = \pm 1$ for up- and down-type initial quarks, respectively.

\subsubsection{Collinear singularities}

Collinear singularities arise only from initial-state photon radiation
in $q\bar{q}$ annihilation.
The collinear part of the $2\rightarrow 3$ cross section is proportional 
to the Born cross section of the hard process with reduced momentum of 
one of the partons. Assuming that parton~$a$ with momentum $p_a$ radiates off a 
photon with $p_{\gamma} = (1-z) p_a$, the parton momentum available for the hard 
process is reduced to $z p_a$. Accordingly, the partonic energy of the total 
process inclusive photon radiation is
        $\tilde{s} = (p_a + p_b)^2 
        = \tilde{\tau} S$\,,
and for the hard process the reduced partonic energy is
        $\hat{s} = (z p_a + p_b)^2 
        = \tau S$\,.
The  'total' and 'hard' variables are thus related by 
        $\hat{s} = z \tilde{s}$ and
        $\tau = z \tilde{\tau}$\,.

Having defined these variables, the partonic cross section in the 
collinear cones can be written in the following way~\cite{Baur:1998kt,Dittmaier:2001ay}
\begin{align}
\begin{split}
        d\hat{\sigma}_{coll} (\hat s) & =\frac{\alpha}{\pi}\,e_q^2\,
                 \int_0^{1-\delta_s}\!\! dz \,\,
                 d\hat{\sigma}_0^{q\bar{q}}(\hat{s})\,\,\kappa_{coll}(z) \quad , \\
\text{with} \qquad
        \kappa_{coll}(z) & = \frac{1}{2} P_{qq}(z) \biggl[
                \ln \biggl( \frac{\tilde{s}}{m_q^2} \, \frac{\delta_{\theta}}{2} \biggr)
                -1 \biggr] + \frac{1}{2} (1-z),
\label{eq_sigma_coll_part}
\end{split}
\end{align}
where $P_{qq}(z) = (1+z^2)/(1-z)$ is an Altarelli-Parisi splitting function 
\cite{Altarelli:1977zs} and $\delta_{\theta}$ is the cut-off parameter to define 
the collinear region by $\cos\theta > 1-\delta_{\theta}$.
The Born cross section refers to the hard scale $\hat{s}$,
whereas in the collinear factor the total energy $\tilde{s}$ is the 
scale needed. In order to avoid an overlap with the soft region, the upper limit 
of the $z$-integration in \eqref{eq_sigma_coll_part} is reduced from $z=1$ 
to $z=1-\delta_s$.

As already mentioned, after adding virtual and real corrections, the mass singularity in  
\eqref{eq_sigma_coll_part} does not cancel and has to be
absorbed into the (anti-)quark density functions. This can be formally achieved by
a redefinition of the parton density functions (PDFs) at NLO QED 
as follows~\cite{Baur:1998kt,Wackeroth:1996hz,Diener:2003ss},
\begin{align}
        f_{a/A}(x)  \rightarrow 
        f_{a/A}(x,\mu_F) \, 
        + &  f_{a/A}(x,\mu_F)\, \frac{\alpha}{\pi}\,e_q^2\, \kappa_{soft}^{PDF}
        + \frac{\alpha}{\pi}\,e_q^2\, \int_x^{1-\delta_s}\! \frac{dz}{z} \,
                 f_{a/A}\Bigl( \frac{x}{z}, \mu_F\Bigr) \,\kappa_{coll}^{PDF}(z) 
\label{eq_renormalizedPDFs}
\\
\text{with} \qquad
        \kappa_{soft}^{PDF} = & - 1 + \ln \delta_s + \ln^2\delta_s 
                - \ln\left(\frac{\mu_F^2}{m_q^2}\right)\,
                        \left[\frac{3}{4} + \ln\delta_s \right] \nn                      
\\
        & + \frac{1}{4} \lambda_{sc} \left[ 9 + \frac{2\pi^2}{3}
                + 3 \ln\delta_s - 2 \ln^2\delta_s \right], \nn
\\[1ex]
        \kappa_{coll}^{PDF}(z) = & \frac{1}{2} P_{qq}(z) \left[
                \ln \biggl( \frac{m_q^2\, (1-z)^2}{\mu_F^2} \biggr) + 1 \right] \nn
\\      
        & - \frac{1}{2} \lambda_{sc} \left[ P_{qq}(z) \ln\frac{1-z}{z}
                        -\frac{3}{2}\frac{1}{1-z} + 2z +3 \right].\nn
\end{align}
The QED factorization scheme dependent $\lambda_{sc}$-parameter is $\lambda_{sc} = 0$
in the $\ms$-scheme and $\lambda_{sc} = 1$ in the $DIS$ scheme.

At the hadronic level, we define the collinear part of the real corrections
for the case where parton $a$ radiates off a collinear 
photon, in the following way by use of~\eqref{eq_renormalizedPDFs},
\begin{align}
\begin{split}
        d\sigma_{coll} (S) = & 
                \frac{\alpha}{\pi}\,e_q^2\, \int\! d\tau\,\int \! \frac{dx}{x}
                 \int_{x}^{1-\delta_s}\!\frac{dz}{z}\,  
                 d\hat{\sigma}_0^{q\bar{q}} (\hat s) \,
                \Bigl[ \kappa_{coll}(z) + \kappa_{coll}^{PDF}(z) \Bigr]
\\
                & \times \biggl[ f_{a/A}        \Big(\frac{x}{z},\mu_F\Bigr)\, 
                         f_{b/B} \Bigl(\frac{\tau}{x},\mu_F\Bigr)\,
                +       f_{b/A} \Big(\frac{\tau}{x},\mu_F\Bigr)\, 
                         f_{a/B} \Bigl(\frac{x}{z},\mu_F\Bigr)\, \biggr],
\label{eq_sigma_coll_had}
\end{split}
\end{align}
where the lower limit of the $z$-integration is constrained to $x$, 
since the parton momentum fraction $x/z$ has to be smaller than unity.
The integral is free of any mass singularity,
\begin{align}
\begin{split}
        \kappa_{coll}(z) +  \kappa_{coll}^{PDF}(z)  &=
                 \frac{1}{2} P_{qq}(z) \, \ln\bigg(
                        \frac{\hat{s}}{z}\frac{(1-z)^2}{\mu_F^2} 
                        \frac{\delta_{\theta}}{2} \bigg)\\
                &+ \frac{1}{2} (1-z) - \frac{1}{2} 
                \lambda_{sc} \left[\, P_{qq}(z) \ln\frac{1-z}{z}
                        -\frac{3}{2}\frac{1}{1-z} + 2z +3 \right].
\end{split}
\end{align}
The $\kappa_{soft}^{PDF}$-term in \eqref{eq_renormalizedPDFs}
cancels the mass singularities owing to soft photons that remain
in the sum of the virtual corrections and 
the soft correction factor $\delta_{soft}^{in}$ in \eqref{eq_delta_soft}.

\subsection{Photon-induced top-squark pair production}

We also consider the photon-induced mechanisms of the top-squark pair production.
At the hadronic level, these processes vanish at leading order owing to the 
non-existence of a photon distribution inside the proton. At NLO in QED, however,
a non-zero photon density arises in the proton as a direct consequence of including
higher order QED effects into the evolution of PDFs, leading thus to non-zero 
photon-induced hadronic contributions.

Feynman diagrams corresponding to the photon--gluon partonic process are 
illustrated in \figref{fig_gy}. Although these are contributions of different
order, they are tree-level contributions to the   
same hadronic final state and thus deserve a closer 
inspection.
The differential cross section for this subprocess is
\begin{align}
\begin{split}
        d\hat \sigma_0^{g\gamma} (\hat s) &=
        \frac{1}{16 \pi \hat s^2}\,\overline{\sum} 
        \bigl\lvert{\cal M}^{g\gamma}_0 (\hat s,\hat{t}_r,\hat{u}_r) \bigr\rvert^2 
        d\hat{t}\; ,
\\
        \overline{\sum} \bigl\lvert{\cal M}_0^{g\gamma}\bigr\rvert^2 & = 
                \frac{1}{4} \cdot \frac{1}{8}\cdot 128 \pi^2 \alpha \alpha_s e_t^2 
                \, N C_F\,
                \left[ 1-2\frac{\hat s \mstopi^2}{\hat{t}_r \,\hat{u}_r} 
                        \left( 1-\frac{\hat s \mstopi^2}{\hat{t}_r \,\hat{u}_r} \right)\right] ,
\label{eq_sigma_phot_ind}
\end{split}
\end{align}
expressed in terms of the reduced Mandelstam variables $\hat{t}_r = \hat{t} - \mstopi^2$, 
 $\hat{u}_r = \hat u -\mstopi^2$.
The quark--photon partonic processes represent contributions of higher order
and we do not include them in our discussion here.

The photon density is part of the
PDFs at NLO QED, which have become available only recently~\cite{Martin:2004dh};
here we present the first study of these effects on the top-squark pair production.

\section{Numerical results}

For the numerical discussion we focus on the production of light top-squark pairs 
$\tilde{t}_1^* \tilde{t}_1$ in proton--proton collisions for LHC energies.
We present the results in terms of the following  hadronic observables:
the integrated cross section, $\sigma$, the differential 
cross section with respect to the (photon inclusive)
invariant mass of the top-squark pair, ($d\sigma/dM_{inv}$), 
the differential cross sections with respect 
to the transverse momentum, ($d\sigma/dp_T$), 
to the rapidity, ($d\sigma/dy$), and to the pseudo-rapidity,
($d\sigma/d\eta$), 
of one of the final state top-squarks.
For getting experimentally more realistic results for the cross sections
we also apply typical sets of kinematical cuts.
A study of the dependence on the various SUSY parameters is given towards 
the end of this section.

The NLO differential cross section at the hadron level is combined from the
contributing partonic cross sections by convolution and summation as follows,
\begin{align}
        d\sigma^{pp}(S) =
                 \int_{\tau_0}^1  \! d\tau \, \left\lbrace \,
                \sum_{i}\frac{d\mathcal{L}_{q_i \bar{q}_i}^{pp}}{d\tau}\,
                         d\hat{\sigma}^{q_i \bar{q}_i} (\hat s)
                + \frac{d\mathcal{L}_{gg}^{pp}}{d\tau} \,
                        d\hat{\sigma}^{gg} (\hat s)
                + \frac{d\mathcal{L}_{g\gamma}^{pp}}{d\tau}\,
                         d\hat{\sigma}_0^{g\gamma} (\hat s) \right\rbrace\,,
\end{align}
where $d\hat{\sigma}^{q_i \bar{q}_i}$ and $d\hat{\sigma}^{gg}$ represent
full one-loop results, including complete virtual and real corrections, and
$d\hat{\sigma}_0^{g\gamma}$ is given in \eqref{eq_sigma_phot_ind}.
The respective parton luminosities refer to~\eqref{eq_luminosity}.

One has to take care of the fact that each
top-squark observed in the laboratory system under a certain angle $\theta$
can originate 
from two different 
constellations at parton level: parton $a$($b$) out of hadron $A$($B$) and vice-versa,
 corresponding to $\theta \rightarrow (\pi - \theta)$.
Both parton level configurations have to be added correctly for hadronic distributions
(for explicit formulas see e.\,g.~\cite{Brein:2007da}).
Note that the two boost factors $\beta$ relating the two partonic center-of-mass (\cm) 
systems with the laboratory system differ by a relative sign, as do the rapidity and the 
pseudo-rapidity of each particle.

Assuming that the forward-scattered parton $a$ carries the momentum fraction $x$ of hadron $A$ and the 
backward-scattered parton $b$ the momentum fraction $\tau/x$ of hadron $B$, the boost factor $\beta$ is given by 
\begin{align}
        \beta &= \frac{x- \tau/x}{x+ \tau/x}\,.
\end{align}
The rapidity of one of the final state top-squarks in the laboratory system, 
$y (\equiv y_{\tilde{t}_1^*})$,
 is related to the rapidity in the partonic \cm frame, 
$y^{cm} = \text{artanh}(p_z^{cm}/E^{cm})$, 
via a Lorentz transformation,
\begin{align}
        y &= y^{cm} - \text{artanh}(-\beta) 
                = y^{cm} + \frac{1}{2} \ln\frac{x^2}{\tau}\,.
\end{align}
The pseudo-rapidity $\eta$ is related to 
$\eta^{cm} = -\ln(\tan\theta^{cm}/2)$ in the c.m.\ frame via
\begin{align}
        \eta &= \text{arsinh} \left( \frac{1}{2} 
                \sqrt{\frac{\mstop^2}{p_T^2} + \cosh^2\eta^{cm}}
                \, \left( \frac{x}{\sqrt{\tau}} - \frac{\sqrt{\tau}}{x} \right)
                + \frac{1}{2} \sinh \eta^{cm} 
                \, \left( \frac{\sqrt{\tau}}{x}+ \frac{x}{\sqrt{\tau}} \right) \right),
\end{align}
which can be derived using the representation
\begin{align}
p = \left( \sqrt{\mstop^2 + p_T^2 \cosh^2 \eta},\,0,\, p_T,\,p_T \sinh\eta\right)
\end{align}
for  the top-squark momentum $p \equiv p_{\tilde{t}_1^*}$.
Since the final state particles are massive, rapidity and pseudo-rapidity 
do not coincide;
in the limit $m \rightarrow 0$ one obtains $\eta = y$.

\subsection{Input parameters and conventions}
\label{subsec_inputs}

Our Standard Model input parameters are chosen in 
correspondance with~\cite{AguilarSaavedra:2005pw},
\begin{gather}
        M_Z = 91.1876\,\text{GeV},\, 
        M_W = 80.403\,\text{GeV},\,
\nn\\
        \alpha^{-1} = 137.036,\, \alpha(M_Z)^{-1} = 127.934,\, 
        G_F = 1.1664 \times 10^{-5}\,\text{GeV}^{-2},\,
\\
        m_t  = 172.7\,\text{GeV},\,
        m_b = 4.7 \,\text{GeV},\,       m_b(m_b) = 4.2 \,\text{GeV}\,. \nn
\end{gather}
All lepton and all other quark masses are set to zero
unless where they are used for regularization.
As a reference we consider the SPA SUSY parameter point
        SPS~1a'~\cite{AguilarSaavedra:2005pw},
unless stated otherwise.
        The current value of the top-quark mass,
        $m_t = 170.9\pm 1.9$\,GeV~\cite{CDFtopmass}, 
        increases the top-squark mass
        $\mstop$ by $0.2\%$, which 
        reduces the total cross section by $\approx 1\%$. 
        The changes for the relative corrections are completely negligible.

For the parton distributions, 
we use the set MRST\,2004\,QED~\cite{Martin:2004dh}, as already mentioned
previously.
Factorization and renormalization scales are chosen equal, 
$\mu_F = \mu_R = 2 m_{\tilde{t}_1}$.

\subsection{Hadronic cross sections and distributions}
\label{subsec_fullCS}

\TABULAR[t]{ccrlcc}{
\hline\hline&&&&\\[-2ex]
scenario &\hspace*{.2cm} channel &
\multicolumn{2}{c}{\hspace*{-1cm}$\sigma^{LO}$ [fb]\hspace*{-1cm}}
&\hspace*{.2cm} $\Delta \sigma^{NLO}$ [fb]\hspace*{.2cm} 
&\hspace*{.2cm}$ \delta = \frac{\Delta \sigma^{NLO}}{\sigma^{LO}}$\\[.5ex]
\hline\hline&&&&&\\[-2ex]
SPS 1a & $\qq$ & 222 &(+0.985)& $-9.71$ &$-4.4 \%$ \\
($m_{\tilde{t}_1} =  376.2\,$GeV)& $gg$ & 1444& & $-15.4$ & $-1.1 \%$ \\
& $\gy$ & &&29.0  & \\[.5ex]
& \bf{total} & \bf{1666} && \bf{3.90} & $\mathbf{0 23\%}$\\[.5ex]
\hline&&&&&\\[-2ex]
SPS 1a' & $\qq$ & 439& (+1.88)& $-11.6$ & $-2.6 \%$ \\
($m_{\tilde{t}_1} =  322.1\,$GeV)& $gg$ & 3292 && $-14.6$ & $-0.44 \%$ \\
& $\gy$ & & & 58.5 & \\[.5ex]
& \bf{total} & \bf{3731} && \bf{32.3} & $\mathbf{0.87 \%}$\\[.5ex]
\hline&&&&&\\[-2ex]
SPS 2 & $\qq$ & 1.17& $(+0.00539)$& $-8.99\times 10^{-2}$ & $-7.7 \%$ \\
($m_{\tilde{t}_1} = 1005.7\,$GeV)& $gg$ & 2.97& & $-3.07\times 10^{-2}$  & $-1.0 \%$ \\
& $\gy$ &&& $15.5\times 10^{-2}$  & \\[.5ex]
& \bf{total} & \bf{4.14} &&$\mathbf{3.44\times 10^{-2}}$& $\mathbf{0.83 \%}$\\[.5ex]
\hline&&&&&\\[-2ex]
SPS 5 & $\qq$ & 2900 &(+10.2) & -13.3  &$-0.46\%$ \\
($m_{\tilde{t}_1} = 203.8 \,$GeV)& $gg$ &31960  && 499 & $1.6\%$ \\
& $\gy$ & & &405 & \\[.5ex]
& \bf{total} & \bf{34860} &&\bf{891}& $\mathbf{2.6\%}$\\[.5ex]
\hline\hline
}
{ Numerical results for the integrated cross sections for 
  light top-squark pair production at the LHC within
  different SPS scenarios 
\protect\cite{AguilarSaavedra:2005pw,Allanach:2002nj,*SPSweiglein}.
\label{tab_results}
}

In~Table~\ref{tab_results} we show results for the cross section for top-squark pair 
production at the LHC within four different scenarios, chosen out of the
SPS benchmark scenarios of the minimal SUGRA 
type \cite{AguilarSaavedra:2005pw, Allanach:2002nj, *SPSweiglein}.
The integrated hadronic cross sections at leading order, $\sigma^{LO}$, 
the absolute size of the EW corrections corresponding to the difference between the LO 
and NLO cross sections, $\Delta\sigma^{NLO}$, and the relative corrections, $\delta$,
given as the ratio of NLO corrections to the respective LO contributions, are presented 
for the $gg$ fusion, the $\qq$ annihilation, and the $\gy$ fusion channel separately. 
The $\gy$ channel contributes only at NLO. 
For the $\qq$ channel, also the numbers for the $\Oas$ pure electroweak Born
level
contributions  
are given in brackets.
These are typically smaller by one order of magnitude compared to the EW NLO corrections.

In scenarios where the top-squark $\tilde{t}_1$ is of intermediate or high mass 
(as SPS~1a, SPS~1a', and SPS~2) 
the NLO contributions are below $1\%$. The corrections to the $\qq$ and 
the $gg$ channels are negative, whereas the 
$\gy$ contribution is always positive and 
of the same size as the other corrections or even larger.
The situation is different in scenarios where the top-squark is very light, 
i.e.~lighter than half of $m_{H^0}$, the
mass of the heavier neutral Higgs boson $H^0$,
where a large fraction of the squarks appears
through production and decay of $H^0$ particles.
This is the case in the SPS 5 scenario [$m_{\tilde{t}_1} = 204\,$GeV,
$m_{H^0} = 694$\,GeV and
$\Gamma(H^0) =  9.7\,$GeV derived from
{\tt FeynHiggs}~\cite{Heinemeyer:1998yj, *Hahn:2006np}]. 
The electroweak contributions in the $gg$ channel are positive
and slightly larger than the $\gy$ fusion contribution.

The interplay of the three production channels is illustrated in 
\figref{fig_allchannels_SPS1aP} where the
absolute EW contributions $\Delta\sigma$ per channel are 
shown as distributions with respect to $\pt$, $M_{inv}$, $y$, or $\eta$.
Owing to the alternating signs, compensations occur where in particular the  
$g\gamma$ channel plays an important role.

\FIGURE[t]{\small
        \pict{0.4}{0}{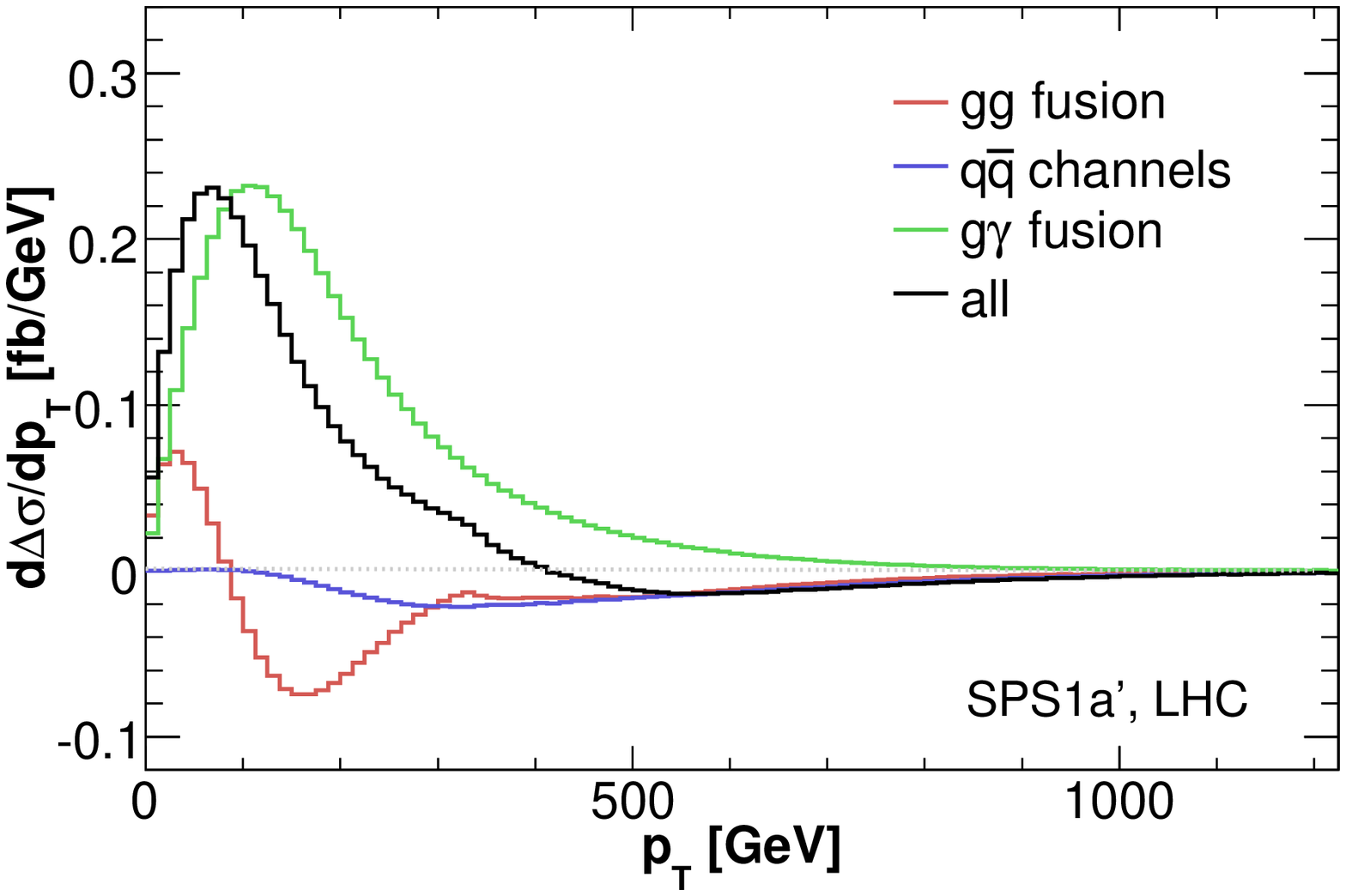}\hspace*{-1ex}%
        \pict{0.4}{0}{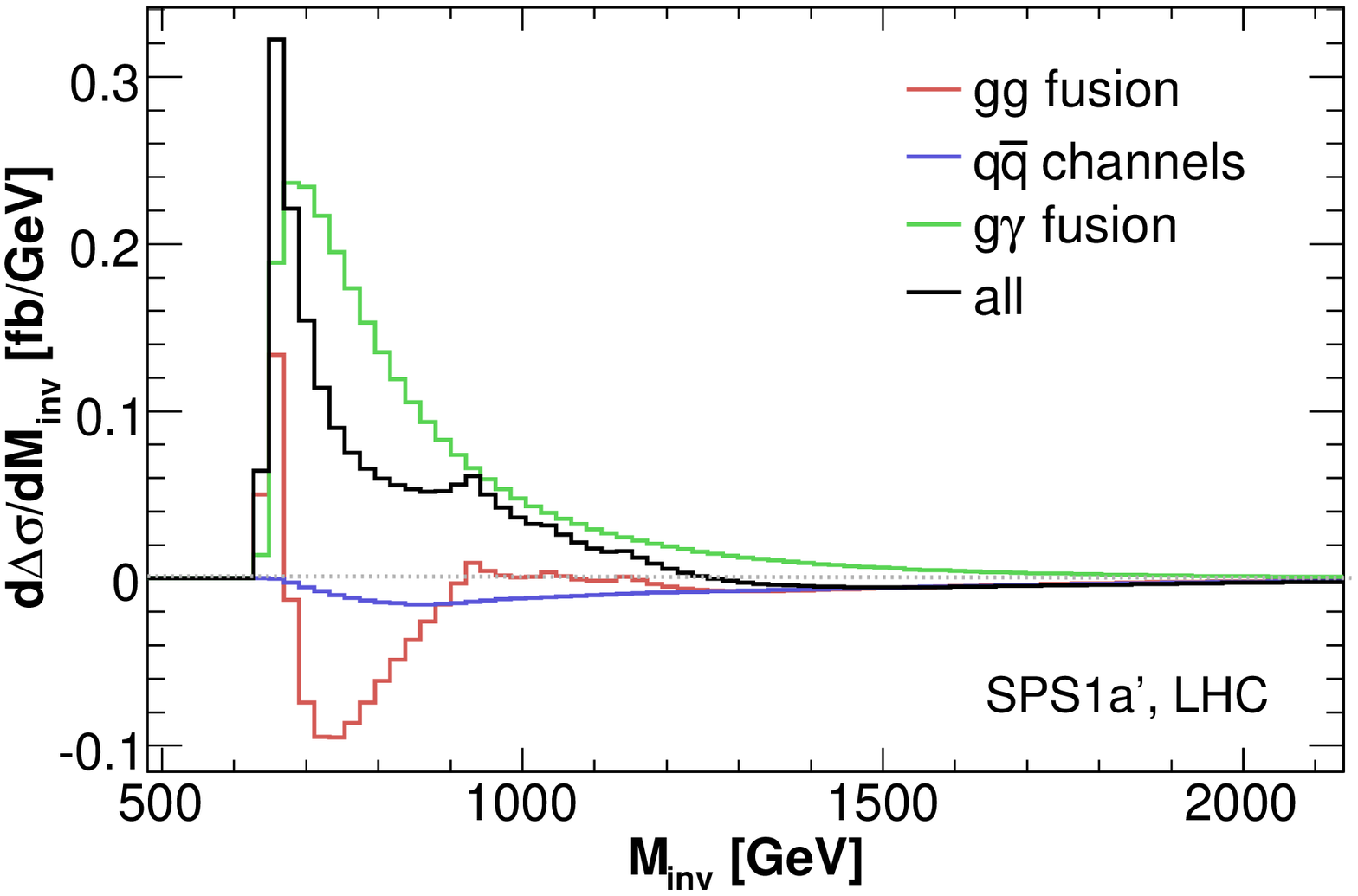}
        \\[1ex]
        \pict{0.4}{0}{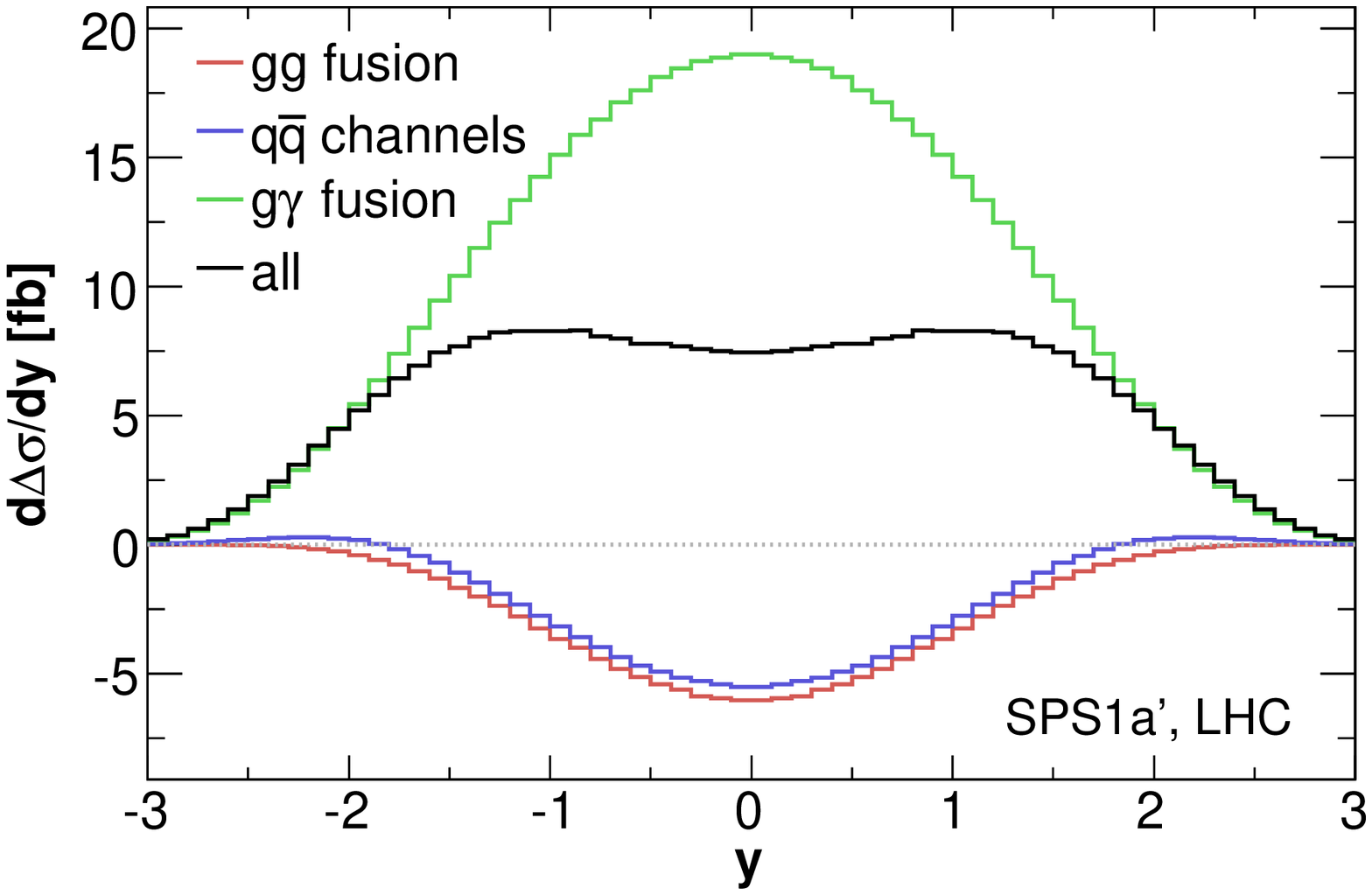}\hspace*{-1ex}%
        \pict{0.4}{0}{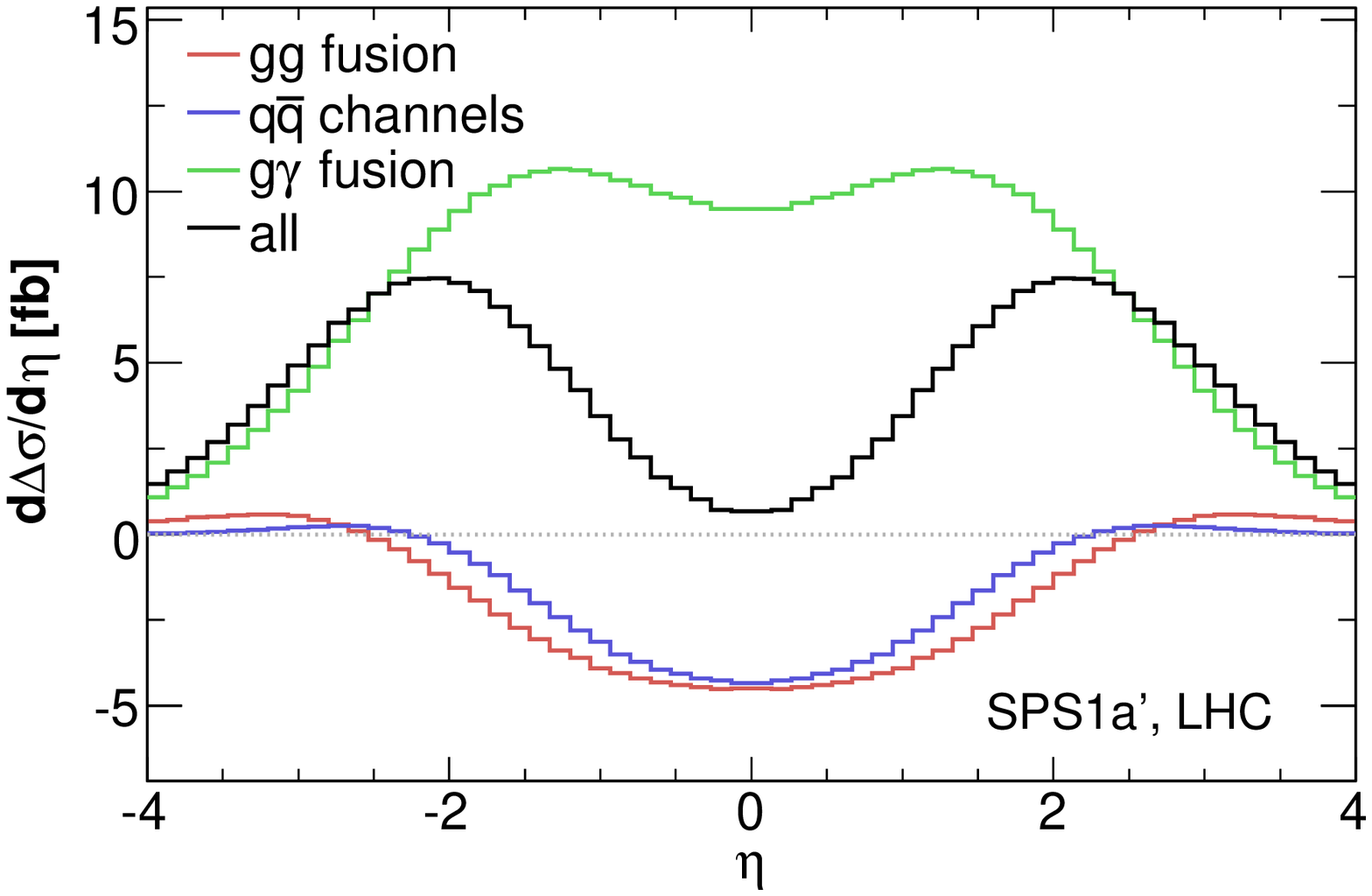}
        \caption{Comparison of EW NLO contributions from the various parton
         channels, for the distributions of 
         transverse momentum $p_T(\tilde{t}_{1}^{\ast})$, 
        invariant mass of the stop pair, rapidity $y(\tilde{t}_1^{\ast})$, 
        and pseudo-rapidity $\eta(\tilde{t}_1^{\ast})$ (from upper left to lower right). 
        $y$ and $\eta$ are given in the laboratory frame.        
        For $gg$ fusion and $\qq$ annihilation, $\Delta$ denotes the difference 
        between NLO and LO
        distributions ($\Delta\sigma \equiv \Delta \sigma^{NLO}$), 
        for $\gy$ one has $\Delta\sigma \equiv \sigma^{\gy}_0$.
        }
        \label{fig_allchannels_SPS1aP}
}

\FIGURE[t]{
        \pict{0.52}{0}{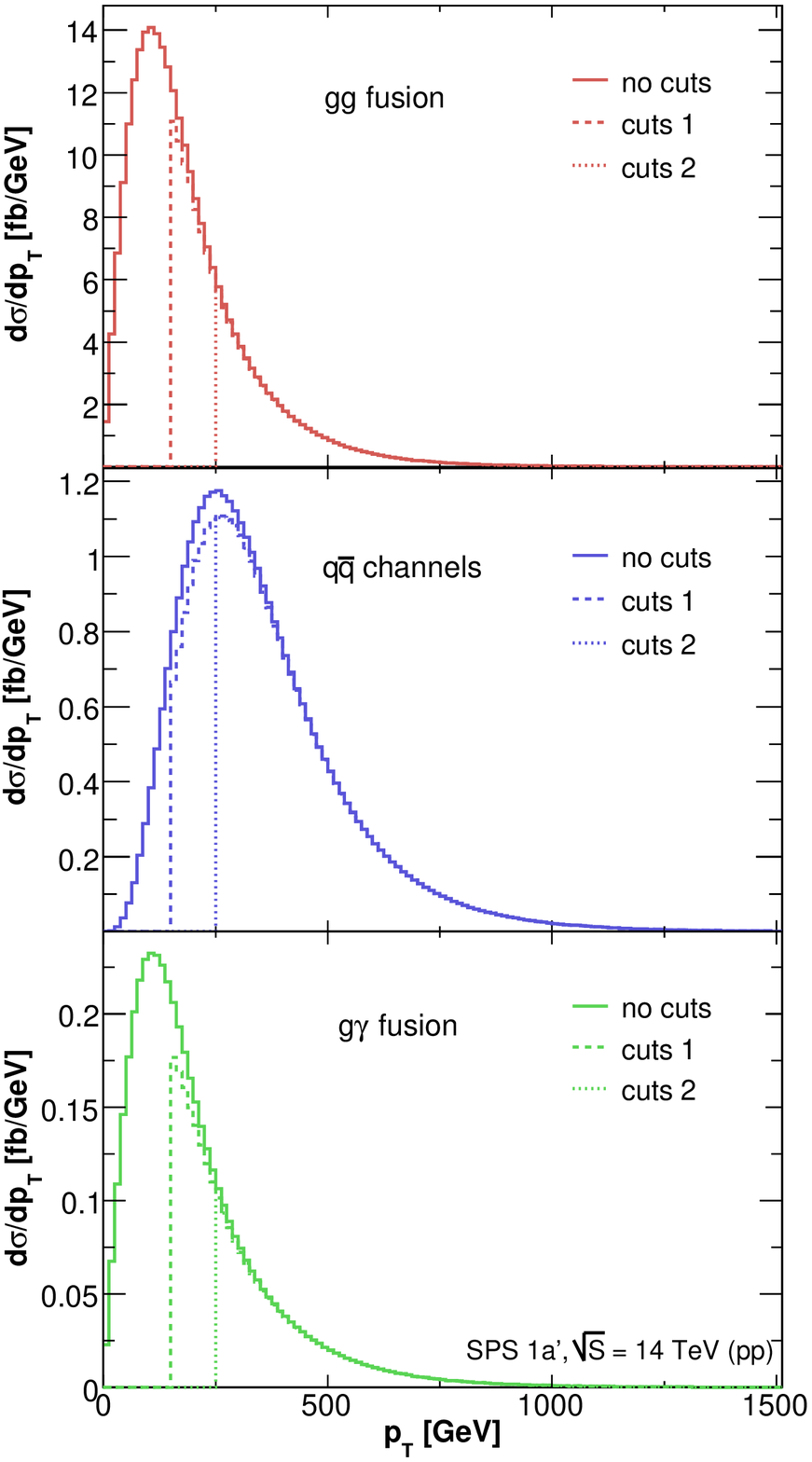}%
        \pict{0.52}{0}{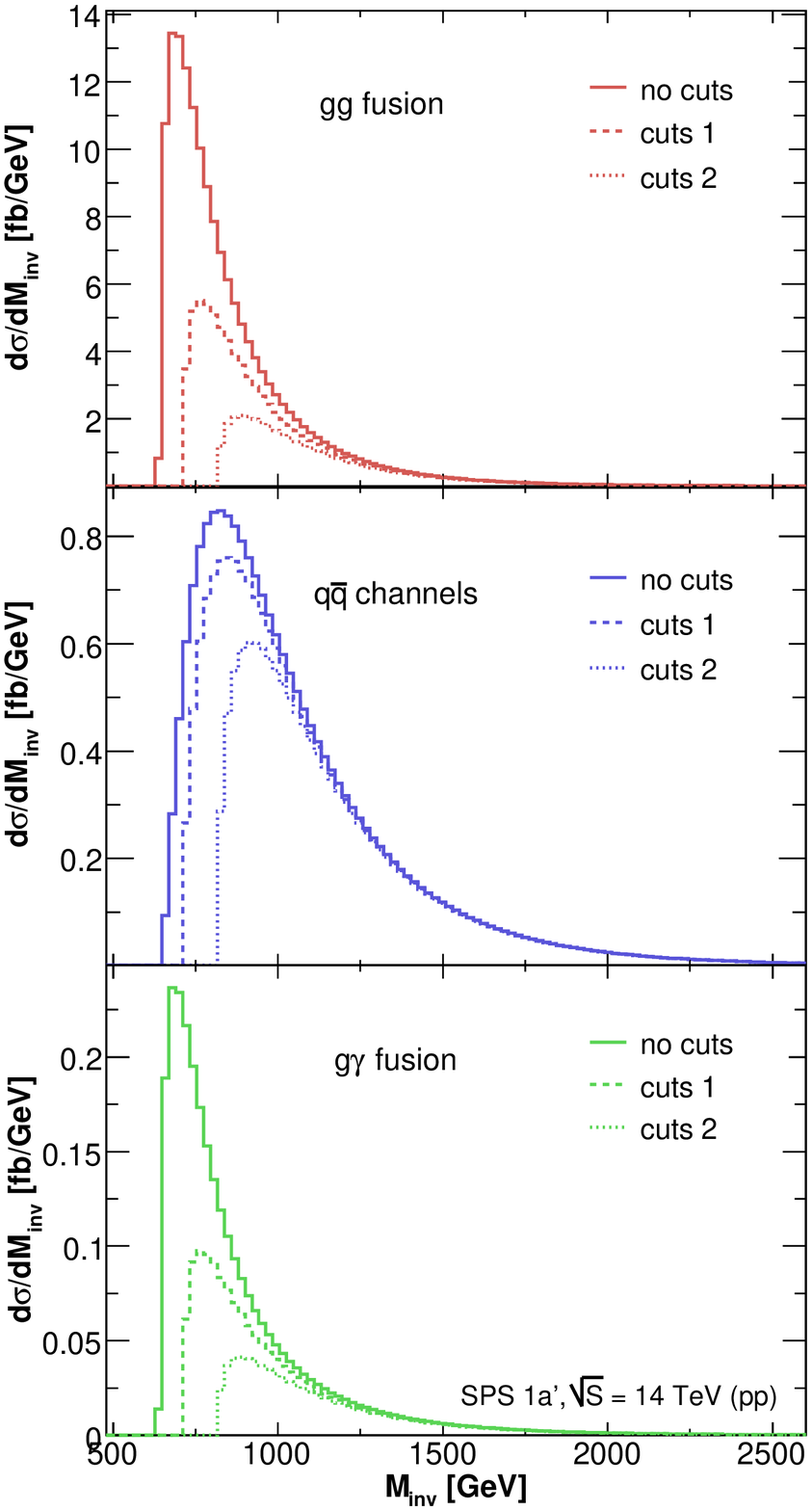}%
        \caption{Comparison of EW NLO differential hadronic cross sections (solid lines) 
        and the distributions where kinematical cuts on the final top-squarks are applied 
        for all three production channels, $gg$ fusion (upper red plots),
        $\qq$ channels (middle blue plots), and $\gy$ fusion (lower green plots). 
        Cuts 1 (dashed lines):  $\mathbf{p_T \ge 150\,}$ {\bf GeV},
        $\mathbf{|\eta| \le 2.5}$,
        cuts 2 (dotted lines):  $\mathbf{p_T \ge 250\,}$ {\bf GeV},
        $\mathbf{|\eta| \le 2.5}$.
         Distributions with respect to the transverse momentum
         $p_T(\tilde{t}_1)$ (left) and the invariant mass of the stop pair (right)
        are shown for $\tilde{t}_1^{\ast} \tilde{t}_1$ pair production at the LHC within the SPS
         1a' scenario.}
        \label{fig_cutdistributions-allchannels}
}

\FIGURE[t]{
        \pict{0.52}{0}{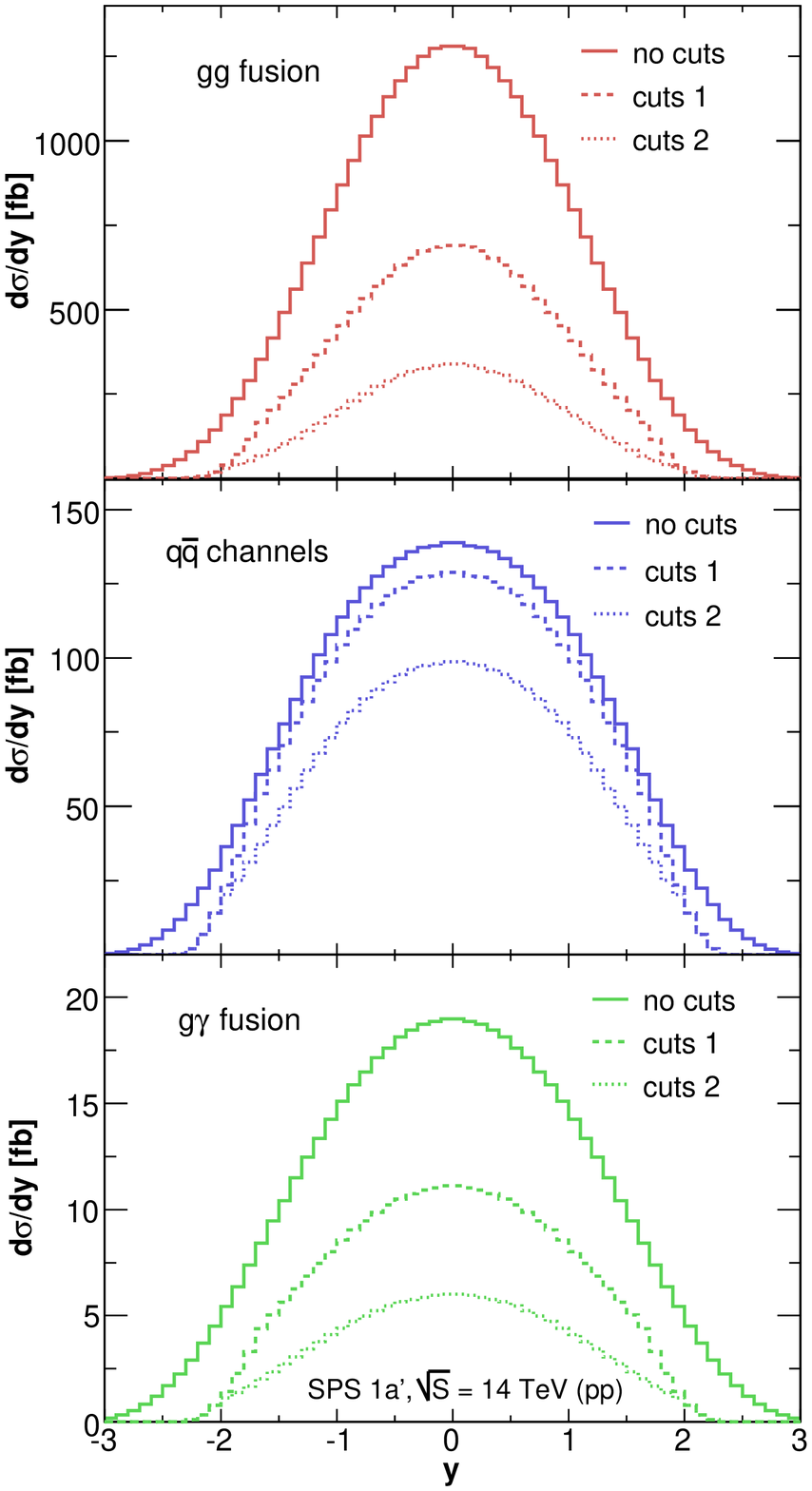}%
        \pict{0.52}{0}{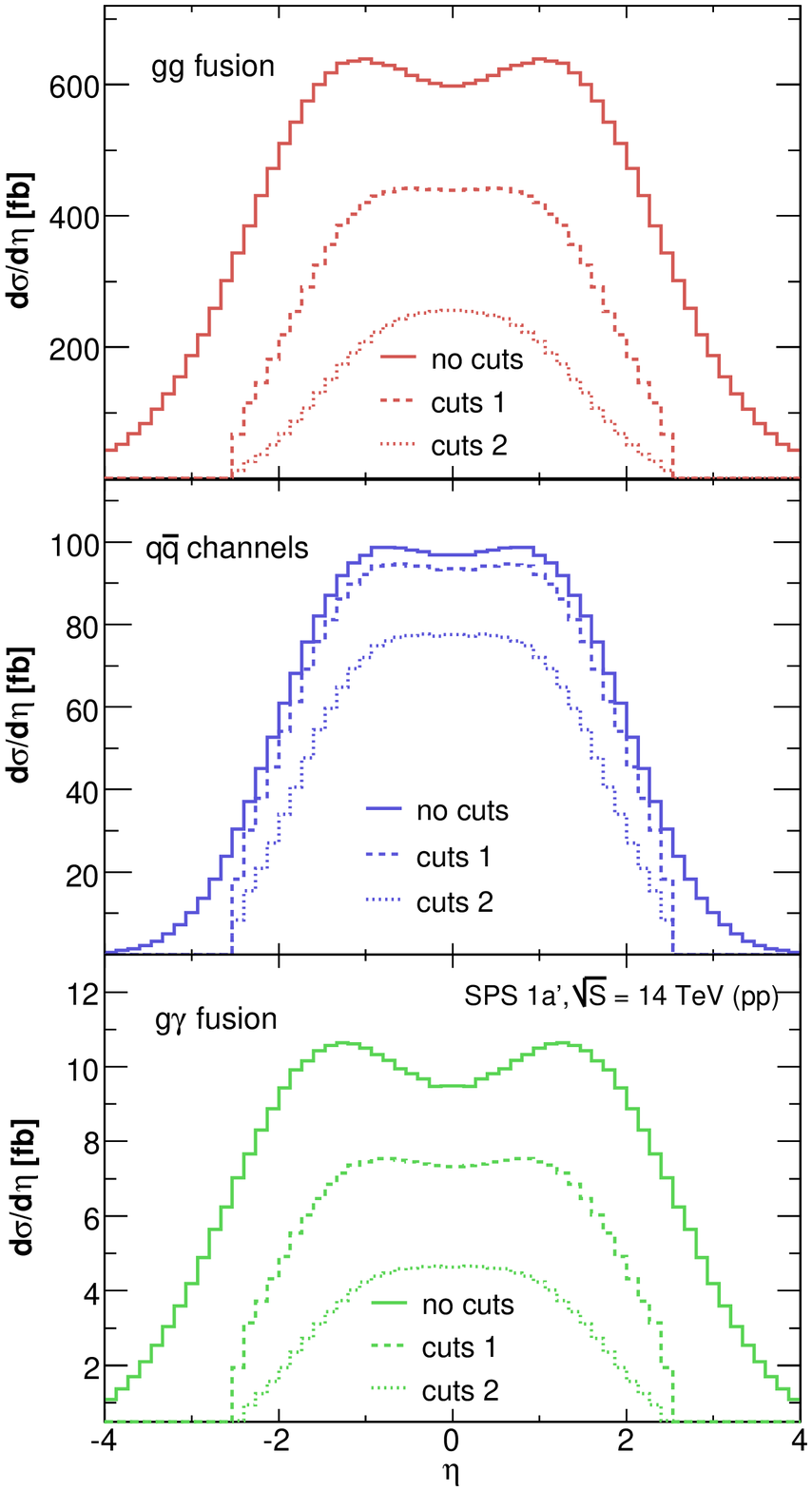}%
        \caption{Same as \figref{fig_cutdistributions-allchannels}, but with respect 
        to the rapidity $y(\tilde{t}_1^{\ast})$ (left) and 
        the pseudo-rapidity $\eta(\tilde{t}_1^{\ast})$ (right).}
        \label{fig_cutdistributions2-allchannels}
}

For realistic experimental analyses, cuts on the kinematically allowed phase
space of the top-squarks have to be applied. 
They can be realized by a lower cut on the transverse momenta of the
final-state particles to focus on high-$\pt$ jets. 
Moreover, detectability of the final state particles 
requires a minimal angle between the particles and 
the beam axis. 
Therefore, we set a cut on the pseudo-rapidity of the top-squarks 
restricting the scattering angle~$\theta$ 
to a central region.
Two exemplary sets of cuts are applied in the following figures
(Figs.~2 -- 5),
\begin{align*}
        \text{cuts 1:\quad} &p_T \ge 150\,\text{GeV and\,\,}  |\eta| \le 2.5 \, 
                \quad\text{(i.\,e.  }9.4^{\circ} \le \theta \le 170.6^{\circ}), \\
        \text{cuts 2:\quad} &p_T \ge 250\,\text{GeV and\,\,}  |\eta| \le 2.5\,.
\end{align*}

The differential cross sections and the influence of cuts are the content of
Figs.~\ref{fig_cutdistributions-allchannels} and~\ref{fig_cutdistributions2-allchannels}. 
Displayed are the hadronic cross sections at NLO, differential  
with respect to $\pt$, $M_{inv}$ and to $y$, $\eta$, respectively.
Both the full (unconstrained) distributions and the
distributions with cuts applied are shown.
The reduction of the integrated cross section owing to the application of cuts
is summarized in Table~\ref{tab_cutwerte}.

\TABULAR{cccc}{
\hline\hline &&&\\[-2ex]
\hspace*{2ex}channel \hspace*{2ex}& \hspace*{2ex}full result\hspace*{2ex} &\hspace*{2ex} $p_T < 150$\,GeV \hspace*{2ex} &\hspace*{2ex} $p_T < 250$\,GeV \hspace*{2ex} \\
 (SPS 1a') & at NLO [fb] & \& $\left|\eta\right| < 2.5$ [fb] & \& $\left|\eta\right| < 2.5$ [fb] \\[.5ex]
\hline&&&\\[-2ex]
$gg$ & 3280 &  1643 ($-50\%$)& 778 ($-76\%$)\\[.5ex]
$\qq$ & 427 & 373 ($-13 \%$) & 280 ($-34 \%$)\\[.5ex]
$\gy$ & 58.5 & 30.6 ($-48 \%$) & 16.2  ($-72 \%$)\\[.5ex]
\hline\hline
}
{Integrated hadronic cross section at NLO within the SPS 1a' scenario for the different production channels.
 Comparison of the full (unconstrained) results and cross sections where cuts on the 
 pseudo-rapidities
$\eta$ and on the transverse momenta $\pt$ of the outgoing top-squarks are applied. 
The relative changes compared to the full results are given in brackets.
\label{tab_cutwerte}}

The application of cuts reduces the $gg$ and $\gy$ channels strongly, 
cutting off the peak of the $\pt$-distributions.  
The reduction is less pronounced 
in the $\qq$ channels where the $\pt$-distribution is  harder. 
The $\pt$-cuts also shift the threshold of the invariant mass distributions 
towards higher values affecting again mainly the $gg$ and $\gy$ channels   
in height and shape. 
The situation for the rapidity distribution is similar.
In the $\qq$ channel, the harder $p_T$-distribution 
goes along with a narrower  $\eta$-distribution,
as shown in the right panels of \figref{fig_cutdistributions2-allchannels}. 
Most of the top-squarks 
produced via $\qq$ annihilation can be found in the central
 region.
In contrast, top-squarks from $gg$ or $\gy$ fusion are often produced 
in the strong forward (or backward) direction, 
and the application of cuts on the pseudo-rapidity thus reduces 
the number of $gg$ or $\gy$ based events significantly.

\medskip

In order to illustrate the numerical impact of the NLO contributions 
on the LO cross section, we show
in \figref{fig_cutdistributions-Kfactor} $K$~factors
$K = \sigma^{NLO}/\sigma^{LO}$  for the $gg$ and the $\qq$ channel, respectively, as 
distributions with respect to $\pt$ and $M_{inv}$. 
The application of cuts influences the $K$~factors only at low values of 
$\pt$ and $M_{inv}$.
The EW corrections in the $\pt$-distribution reach typically $-10\%$ in the $gg$ channel, 
and  $-20\%$ in the $\qq$ channel, for large values of $\pt$.
In the invariant mass distributions, they are somewhat smaller, but still
sizeable, at the 10\% level for large $M_{inv}$.
The large effects at high $\pt$ and $M_{inv}$ are dominated by the 
double logarithmic contributions arising from virtual $W$ and $Z$ bosons in
loop diagrams.

The small peaks visible in the $gg$ invariant mass distribution
correspond to two-particle thresholds related to 
$\tilde{b}_1^{\ast} \tilde{b}_1$, $\tilde{b}_2^{\ast} \tilde{b}_2$, and 
$\tilde{t}_2^{\ast} \tilde{t}_2$ pairs in $gg$ vertex and box diagrams,  
illustrated in~\figref{fig_gg-virtualcorrs} 
[in the SPS~1a' scenario, the masses of the involved squarks are 
        $m_{\tilde{b}_1} = 460.7\,$GeV, 
        $m_{\tilde{b}_2} = 514.8\,$GeV, 
        $m_{\tilde{t}_2} = 569.4\,$GeV].
Thresholds from the squarks of the first two generations are CKM suppressed.
The threshold effects appear also in the 
$\pt$-distribution, around $300\,$GeV, 
but they are smeared out and much less pronounced.

\figref{fig_cutdistributions-Kfactortot} shows total $K$~factors, defined as
$K = (\sigma^{NLO}_{gg} +\sigma^{NLO}_{\qq}+ \sigma^{LO}_{\gy}) / (\sigma^{LO}_{gg} + \sigma^{LO}_{\qq} )$.  
It is obvious that, although small for the total cross section, the EW higher order 
contributions cannot be neglected for differential distributions where, 
in the high-$\pt$ and 
high-$M_{inv}$ range, they are of the same order of 
magnitude as the SUSY-QCD corrections~\cite{Beenakker:1994an,*Beenakker:1996ch}.

\FIGURE[tb]{
        \pict{0.4}{0}{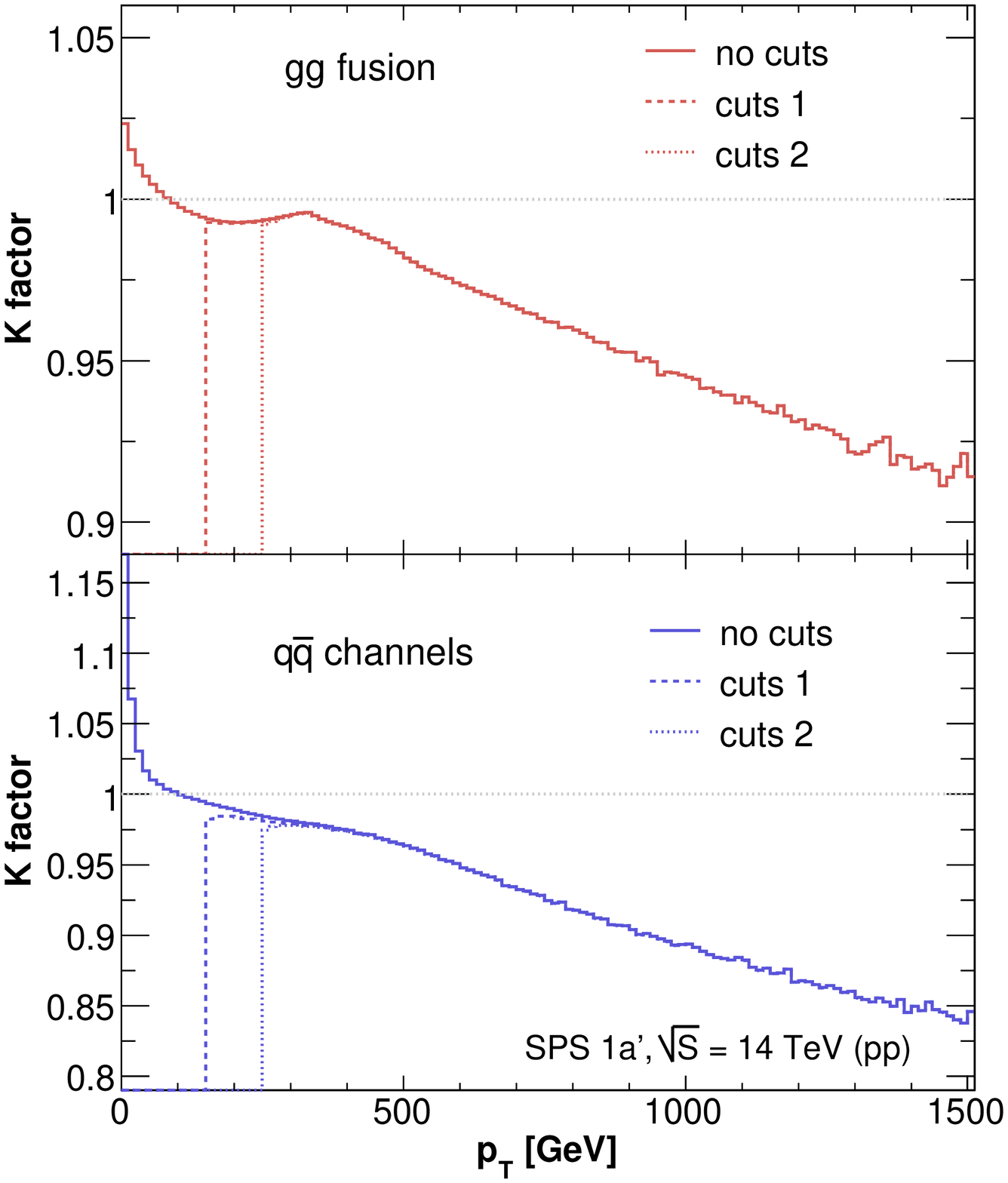}%
        \pict{0.4}{0}{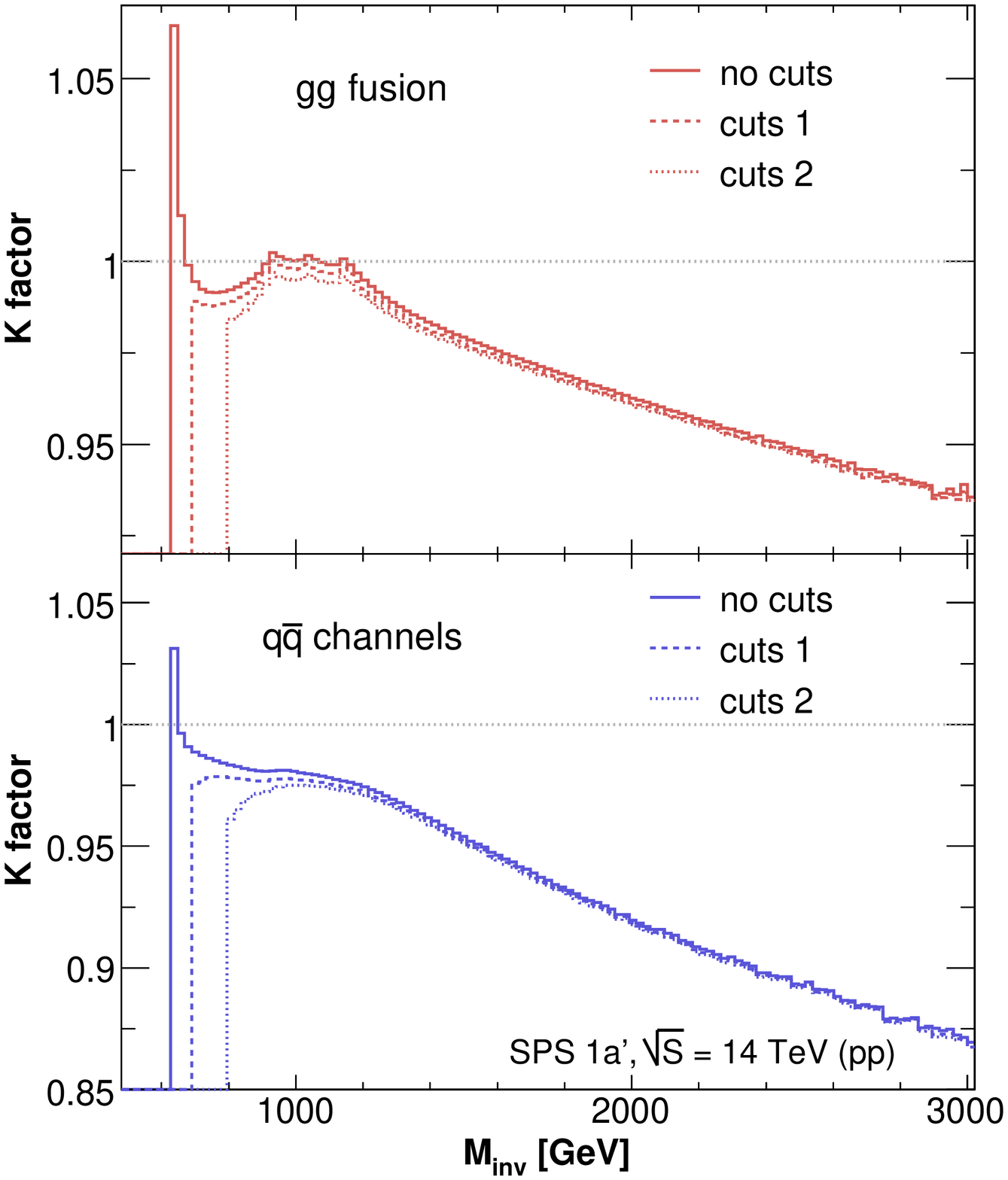}%
        \caption{Same as \figref{fig_cutdistributions-allchannels}, but shown are 
        the $K$ factors, $K = \sigma^{NLO} / \sigma^{LO}$, 
        for $gg$ fusion (upper plots) and $\qq$ channels (lower plots).}
        \label{fig_cutdistributions-Kfactor}
}

\clearpage

\FIGURE[t]{
        \pict{0.4}{0}{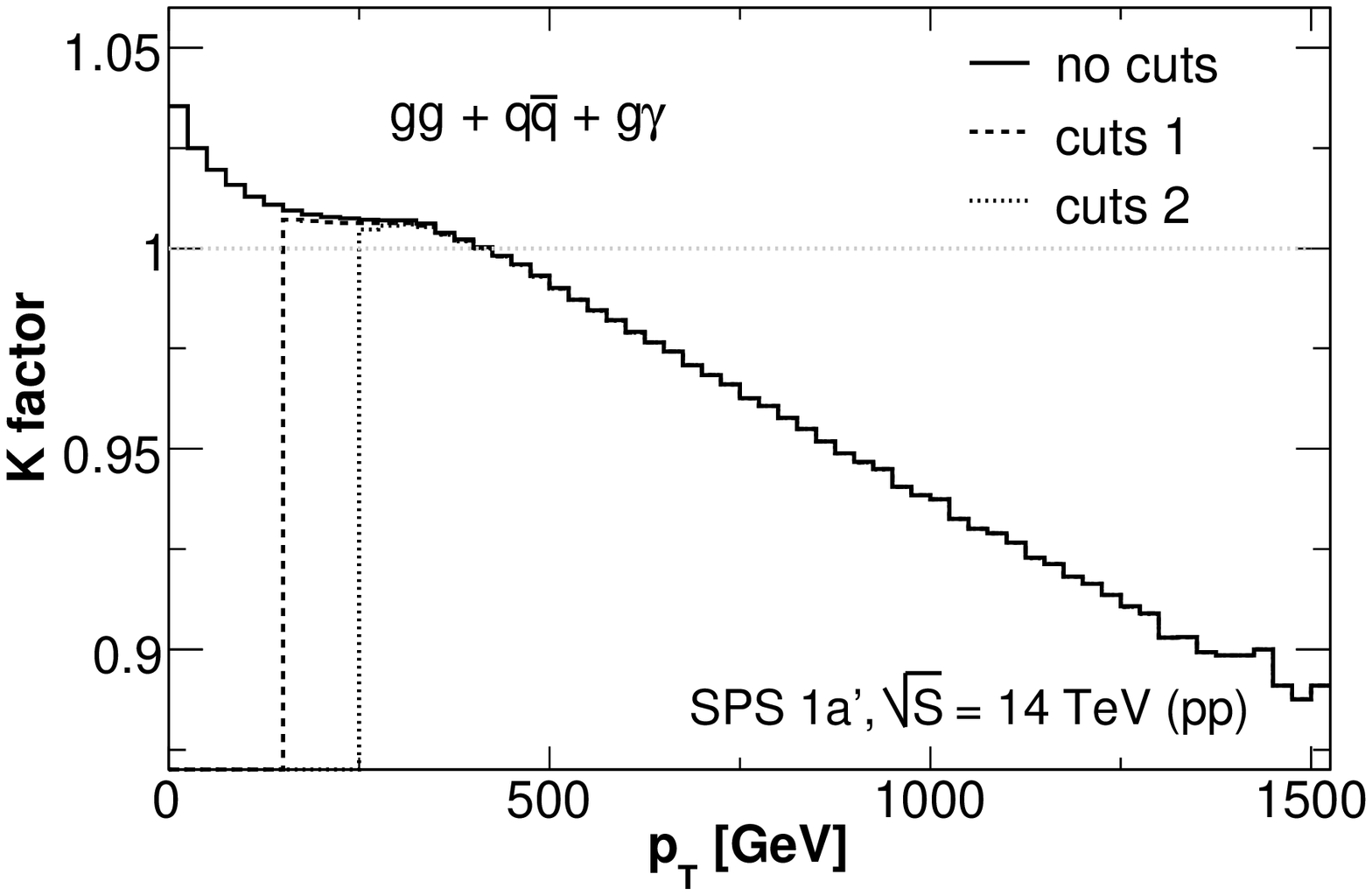}%
        \pict{0.4}{0}{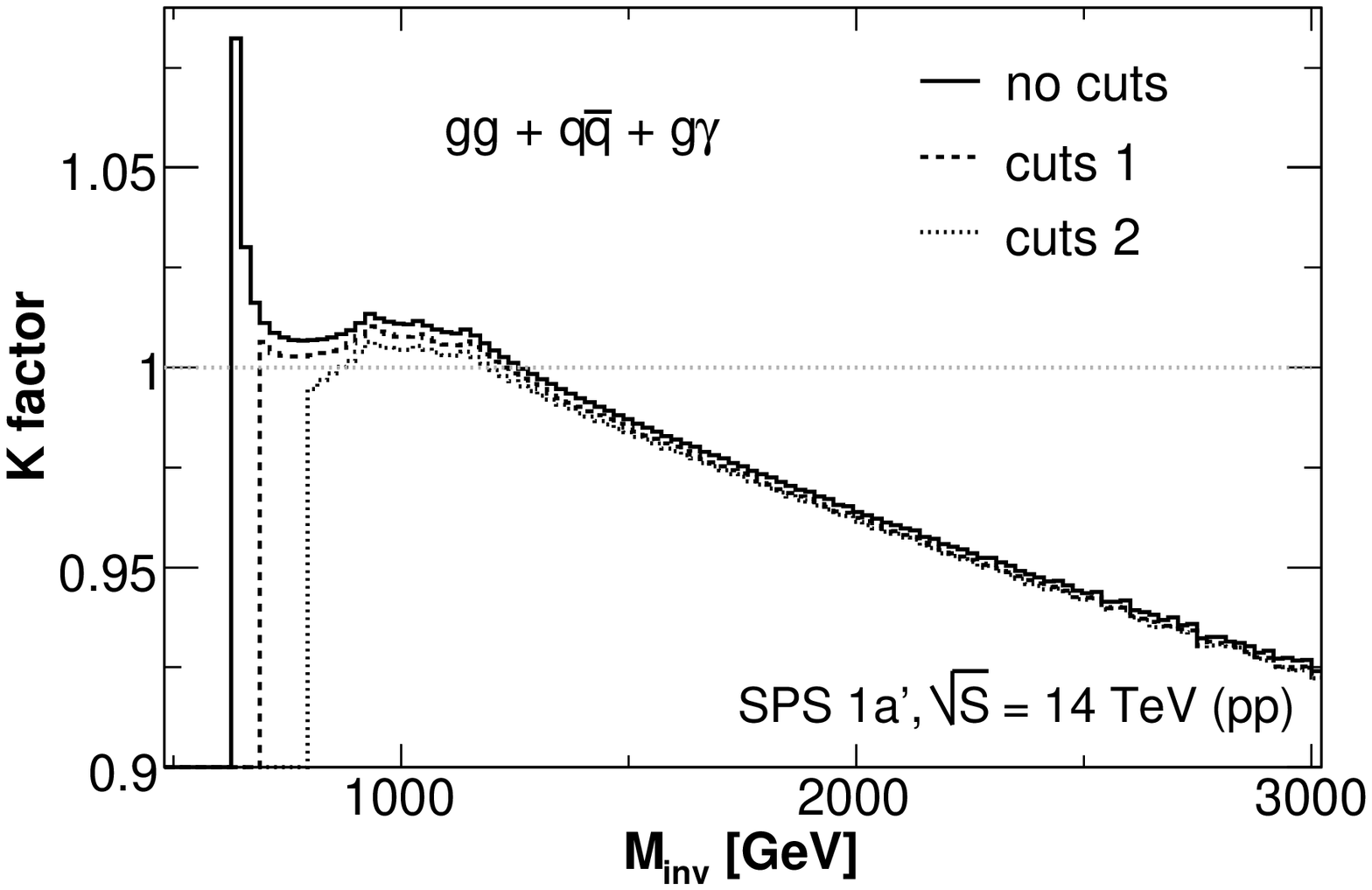}%
        \caption{Same as \figref{fig_cutdistributions-allchannels}, but shown is 
        the total $K$ factor,
$K = (\sigma^{NLO}_{gg} +\sigma^{NLO}_{\qq}+ \sigma^{LO}_{\gy}) / ( \sigma^{LO}_{gg} + \sigma^{LO}_{\qq} )$.}
        \label{fig_cutdistributions-Kfactortot}
}


\subsection{SUSY parameter dependence}
\label{subsec_susypars}

In order to study the dependence of the EW contributions on 
the various SUSY parameters in more detail,
we consider the ratio of the NLO contribution in each channel 
to the combined $gg + \qq$ Born cross section,
 $\delta_{tot} = \Delta\sigma^{NLO}_{\{gg,\, \qq,\, \gy\}} / \sigma^{LO}_{tot}$.
We focus on those parameters that determine the top-squark mass, cf. \eqref{eq_stopmass},
and vary each quantity out of the set $\msq$, $\msu$, $\tb$, $A_t$, or $\mu$ 
around its SPS~1a' value while keeping all other parameters fixed to those 
of the default SPS~1a' scenario.
The results are displayed in the left panels of 
Figs.\,\ref{fig_MSQ3_dep}~--~\ref{fig_Mu_dep}. 
Simultaneously, we show  the mass of the light top-squark $\tilde{t}_1$
as a function of the varied parameter in the respective right panels (black solid lines). 
The parameter configuration
 of the SPS~1a' scenario is marked by a vertical gray dotted line in all the figures.

\FIGURE[tb]{
        \pict{0.4}{0}{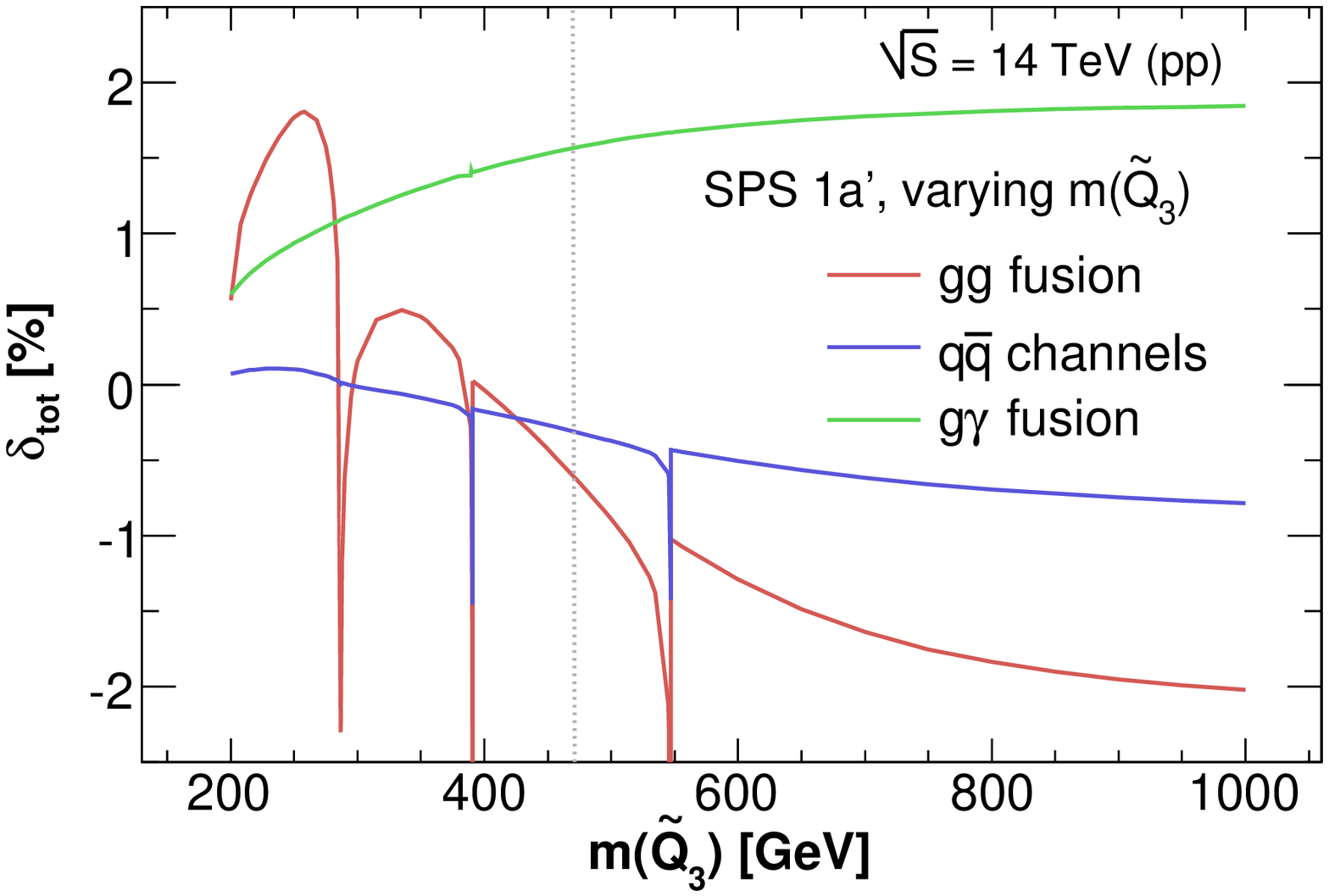}%
        \pict{0.4}{0}{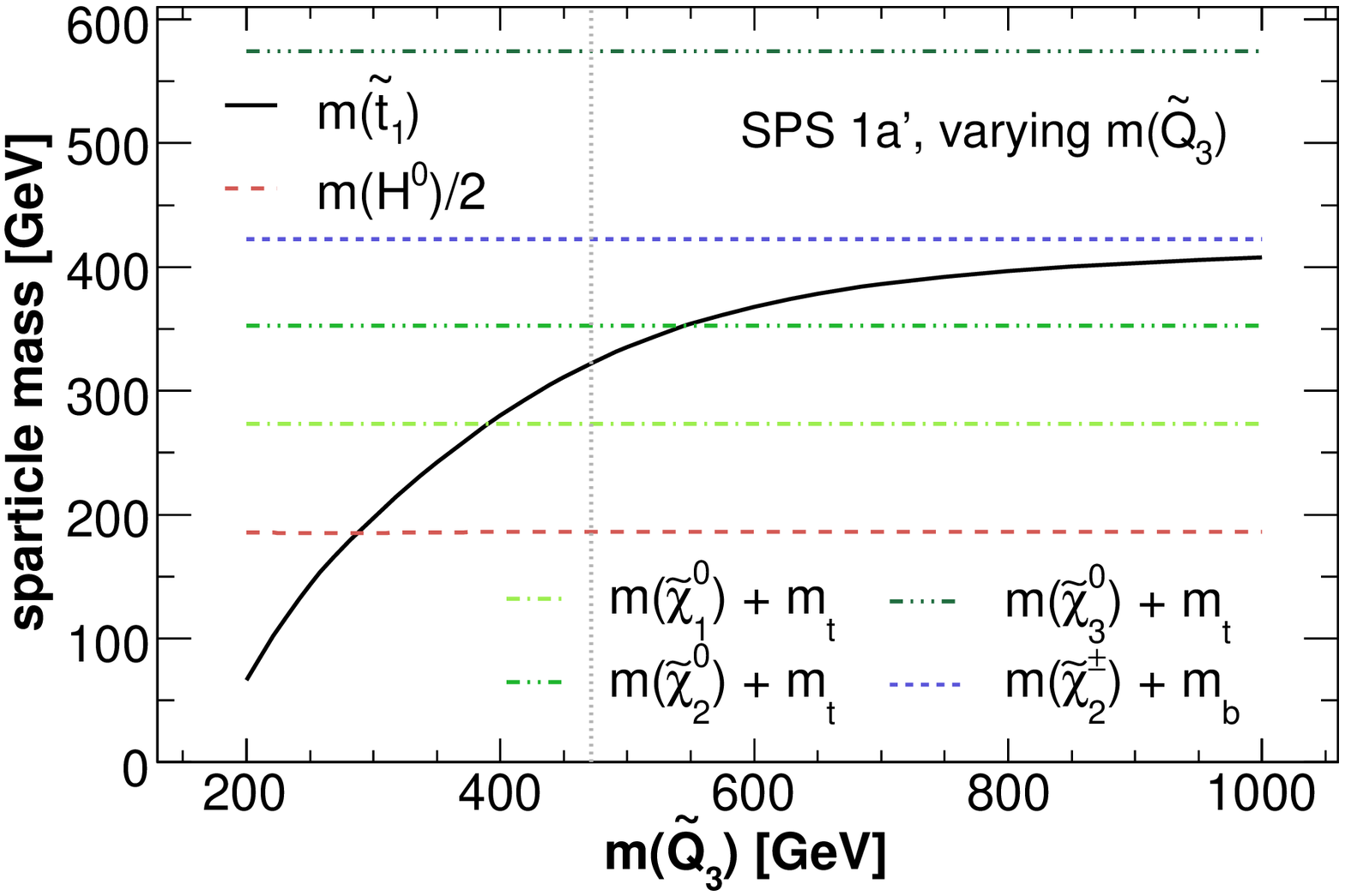}%
        \caption{Left: Relative EW corrections as a function
         of the soft-breaking parameter $\msq$ for each of the
         indicated channels compared to the combined ($gg + \qq$) LO cross section in
         the SPS 1a' scenario where $\msq$ is varied around 
         the SPS1a' value (gray dotted line).
        Right: Mass of $\tilde{t}_1$, half of the mass of $H^0$, sums of the masses of 
        the top-quark and $\tilde{\chi}_1^0$, $\tilde{\chi}_2^0$, and $\tilde{\chi}_3^0$,       
        respectively, and sum of the masses of the bottom-quark and $\tilde{\chi}_2^{\pm}$
        as a function of $\msq$. 
        All other parameters are chosen according to the SPS1a' scenario.}
        \label{fig_MSQ3_dep}
}

\FIGURE[t]{
        \pict{0.4}{0}{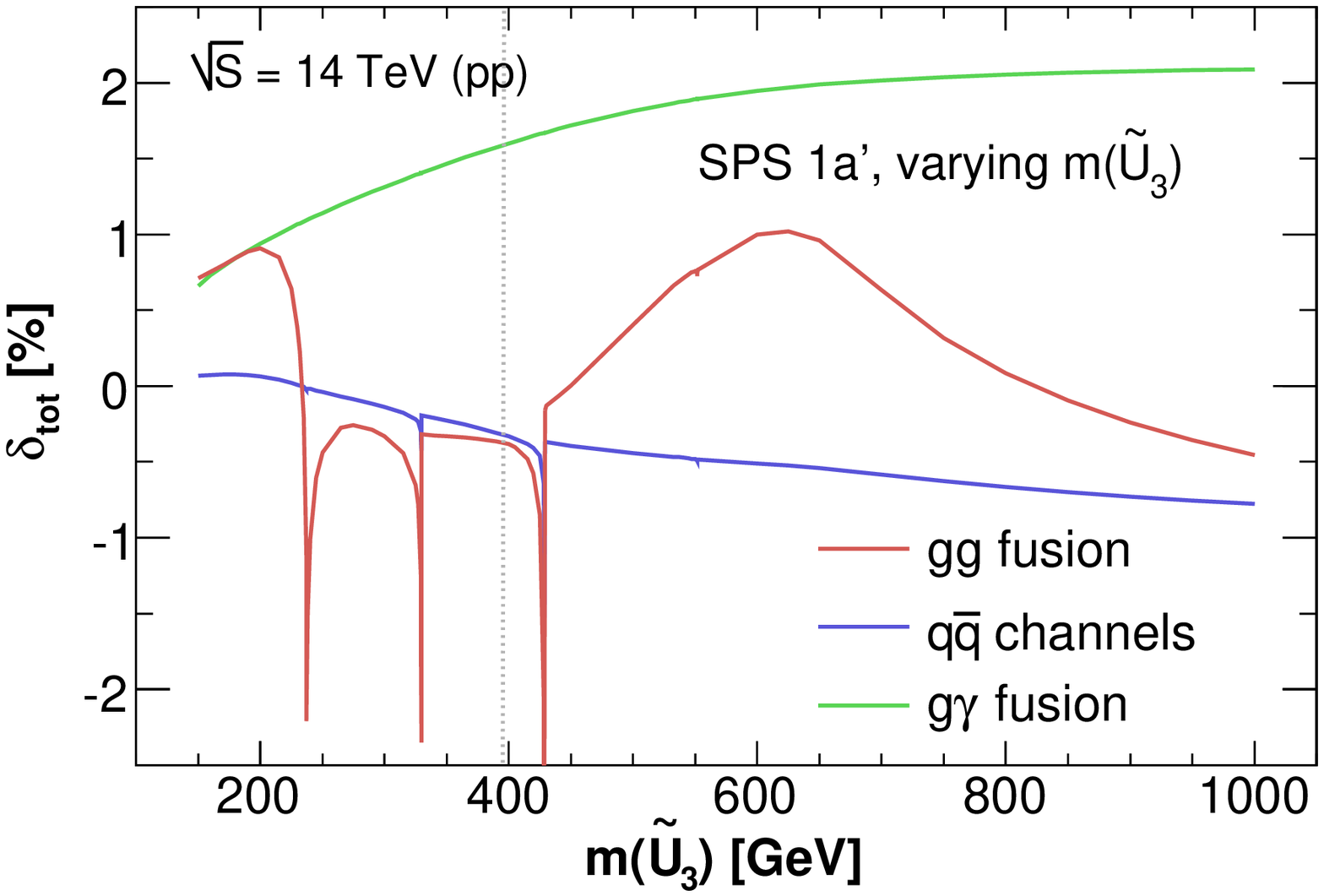}%
        \pict{0.4}{0}{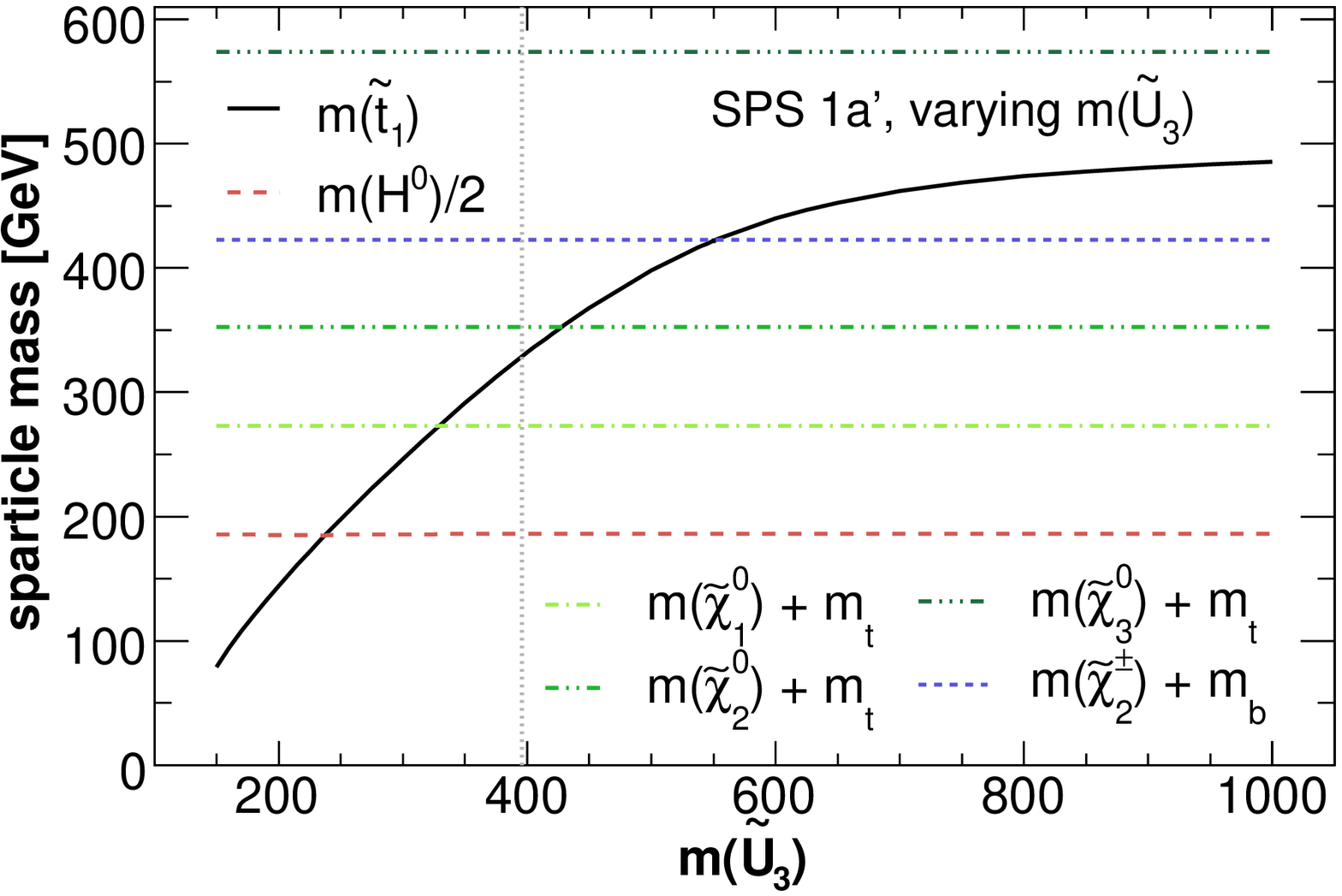}%
        \caption{Same as \figref{fig_MSQ3_dep}, but for variation of the 
        soft-breaking parameter $\msu$.}
        \label{fig_MSU3_dep}
}

\FIGURE[t]{
        \pict{0.4}{0}{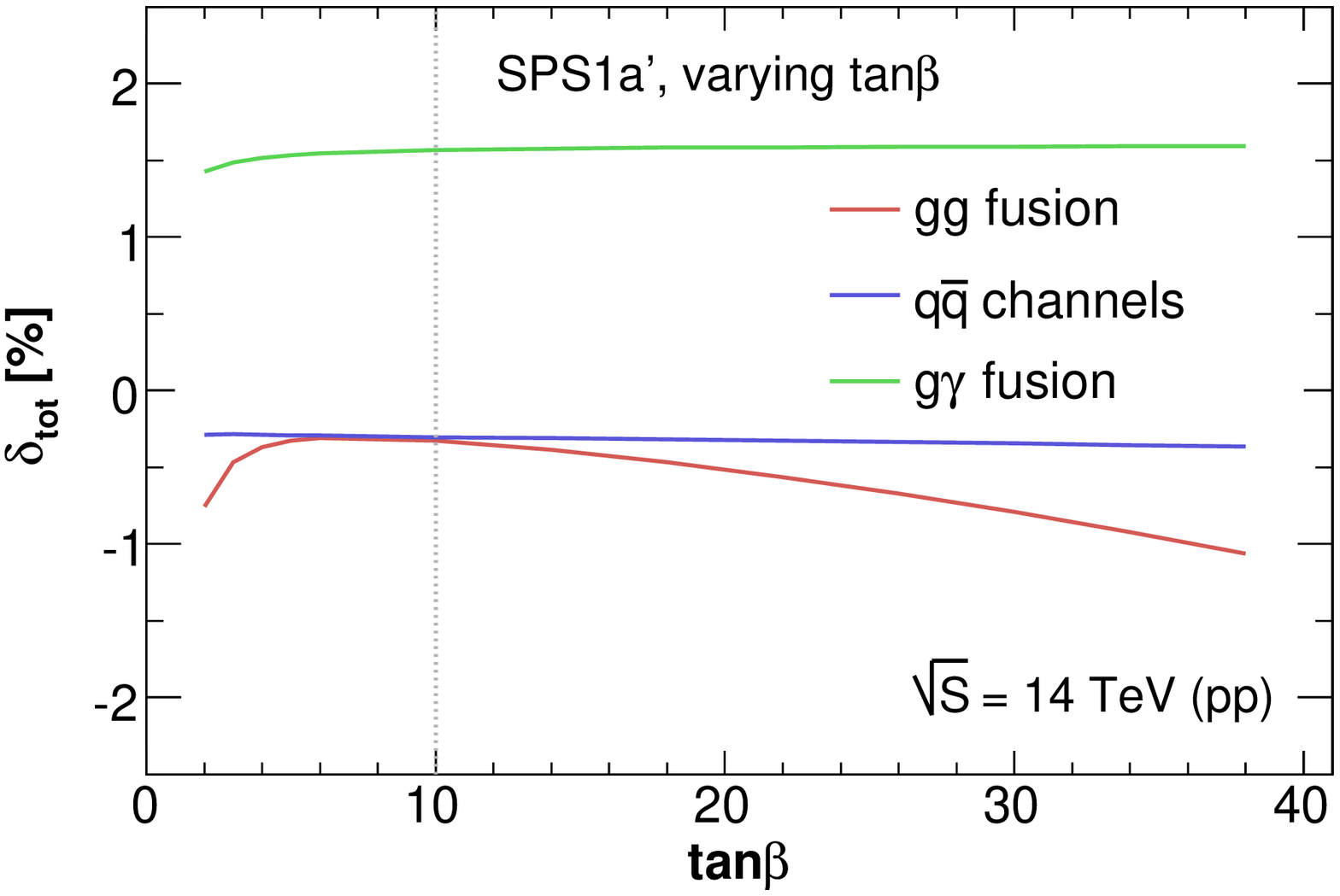}%
        \pict{0.4}{0}{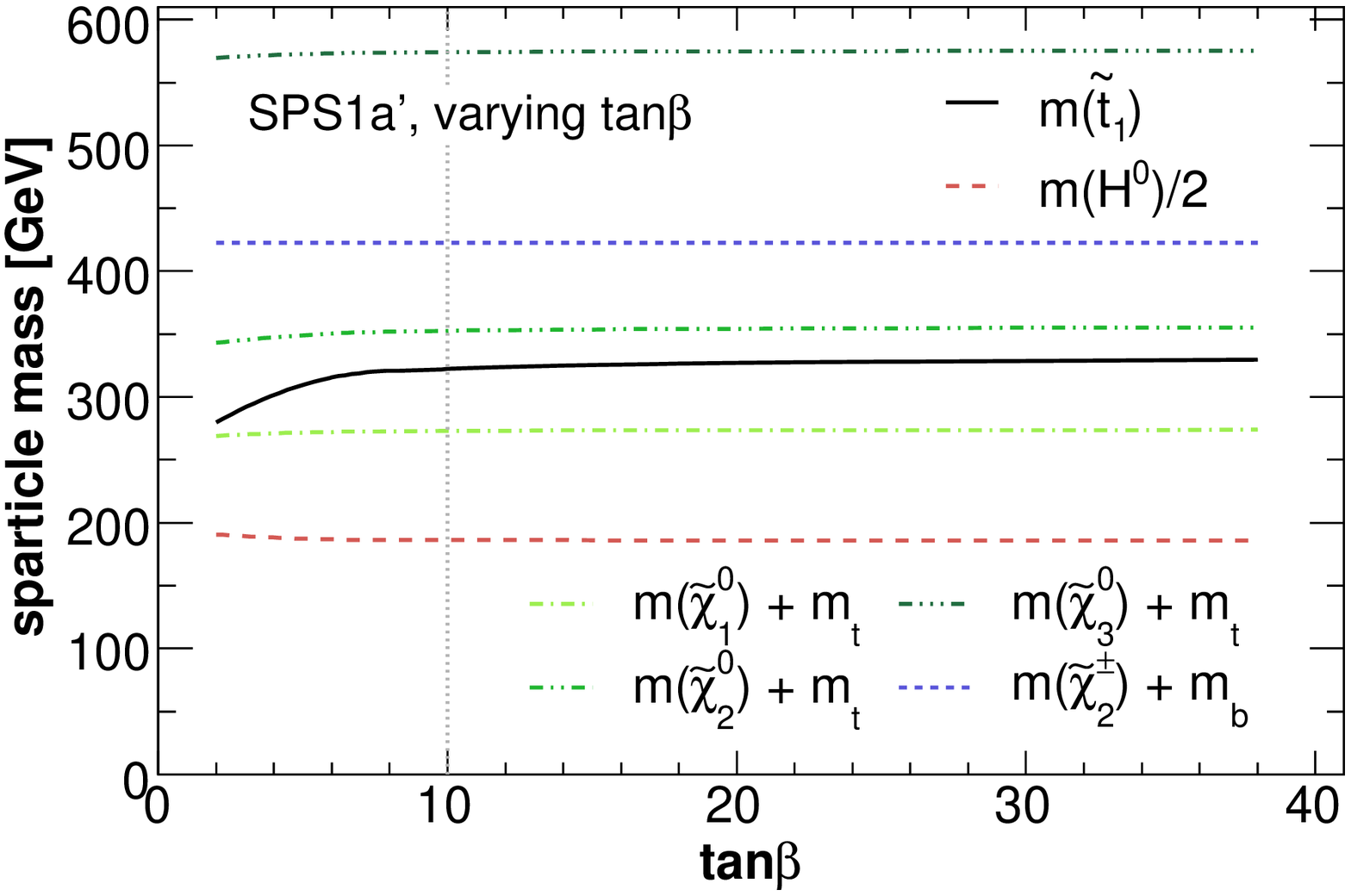}%
        \caption{Same as \figref{fig_MSQ3_dep}, but for variation of $\tb$.}
        \label{fig_tanbeta_dep}
}

We find the following general behaviors. 
The $\gy$ contributions are from tree level diagrams and the only relevant parameter 
is thus the top-squark mass $\mstop$. 
In all scenarios, the $\gy$ fusion channel is as important as the EW corrections 
to the $\qq$ and $gg$ processes.
The $\qq$ corrections, being practically always negative,
involve many different SUSY particles in the loops,
although the relative corrections show only small variations.
The $gg$ contributions are more sensitive to the considered SUSY
parameters. 
The plots show  striking peaks (some of them are also visible in 
$\qq$ annihilation), which correspond to threshold effects
and can be explained by the SUSY particle masses 
displayed at the right panels of Figs.\,\ref{fig_MSQ3_dep} --
\ref{fig_Mu_dep}. 
They occur in the Higgs-exchange diagrams when
$\mstop = m_{H^0}/2$ (red long-dashed lines in the figures), 
and in the top-squark wave function renormalization 
when $\mstop$ equals the sum of masses of a neutralino and the top-quark 
(green dash-dotted lines) or of a chargino and the bottom-quark 
(blue dashed lines). 
The chargino-induced peaks are less pronounced than those from neutralinos 
and not visible in  \figref{fig_MSU3_dep} and \figref{fig_Mu_dep}.

Outside of such singular parameter configurations,  
over a wide range of SUSY parameters, 
the combined EW contributions to top-squark pair production 
are only weakly parameter dependent.

\FIGURE[t]{
        \pict{0.4}{0}{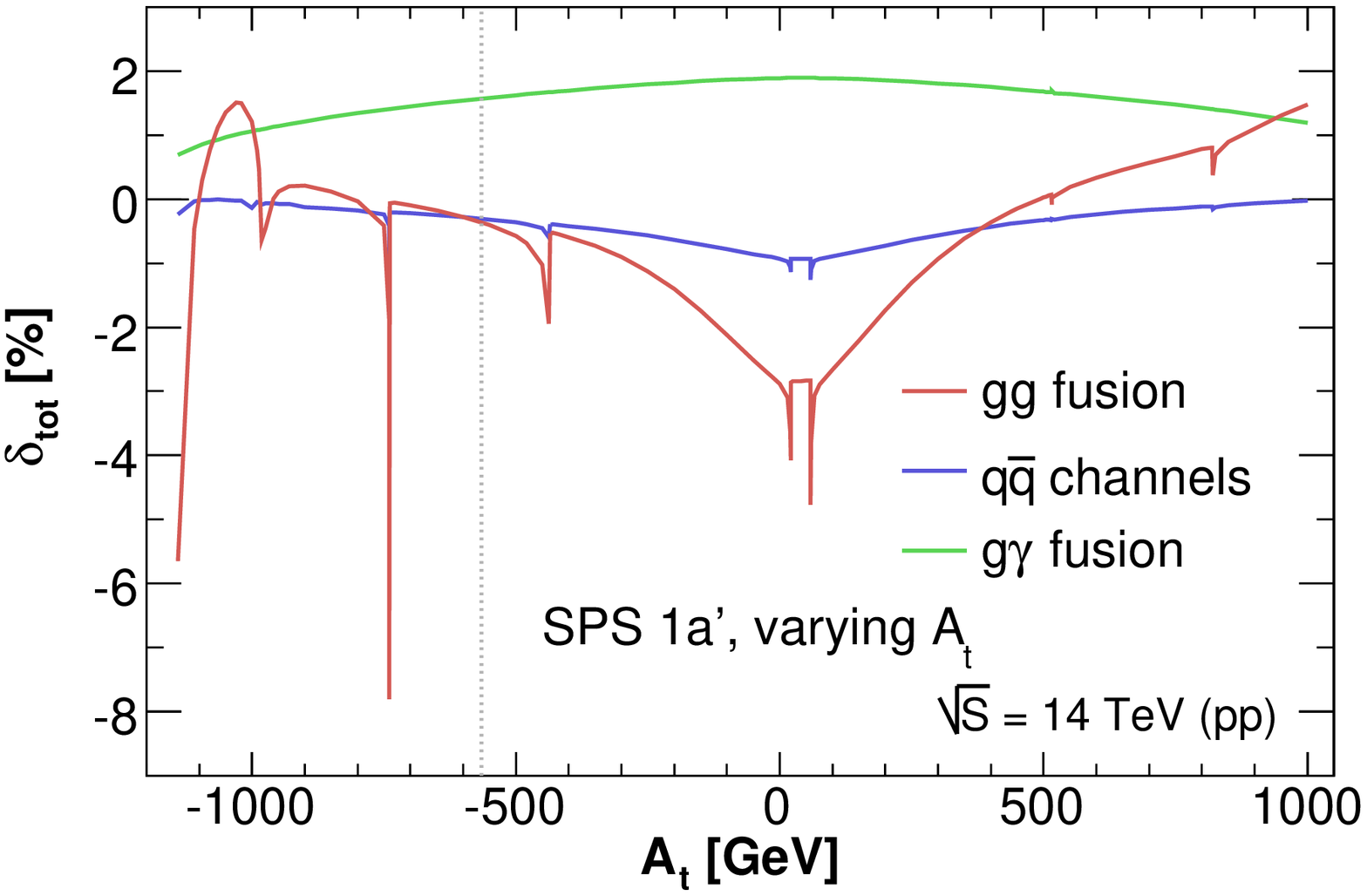}%
        \pict{0.4}{0}{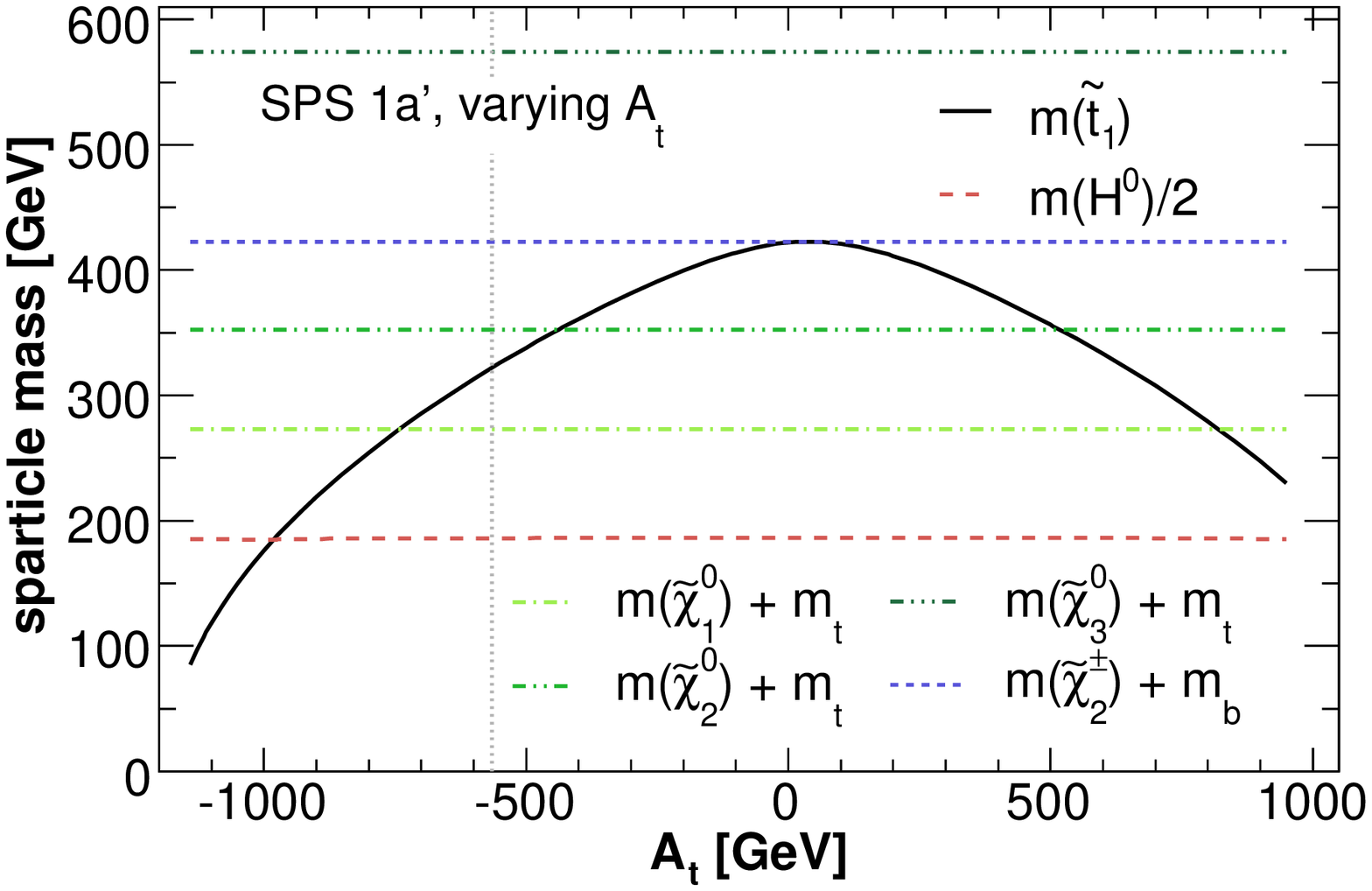}%
        \caption{Same as \figref{fig_MSU3_dep}, but for variation of trilinear coupling parameter $A_t$.}        
        \label{fig_At_dep}
}

\FIGURE[t]{
        \pict{0.4}{0}{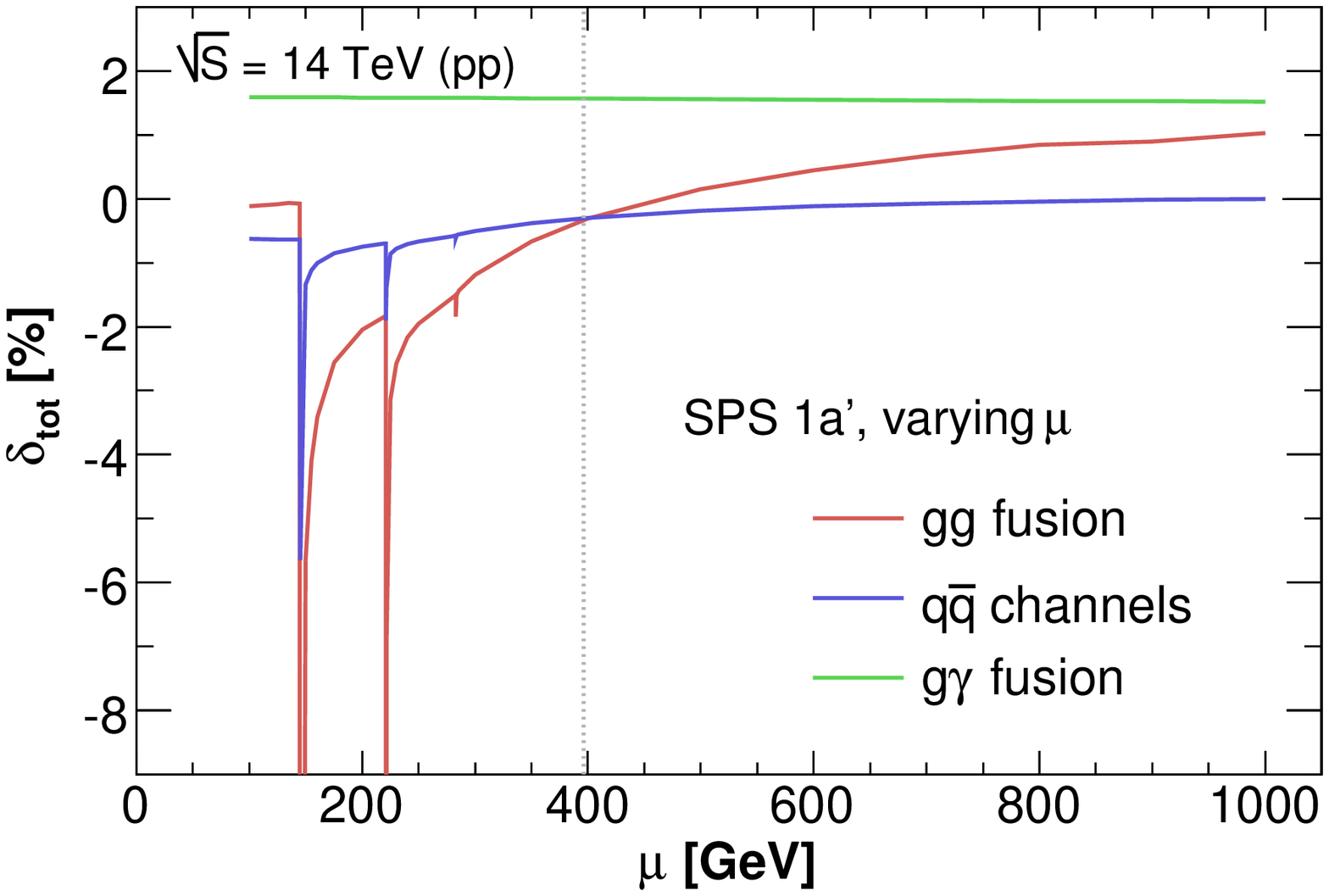}%
        \pict{0.4}{0}{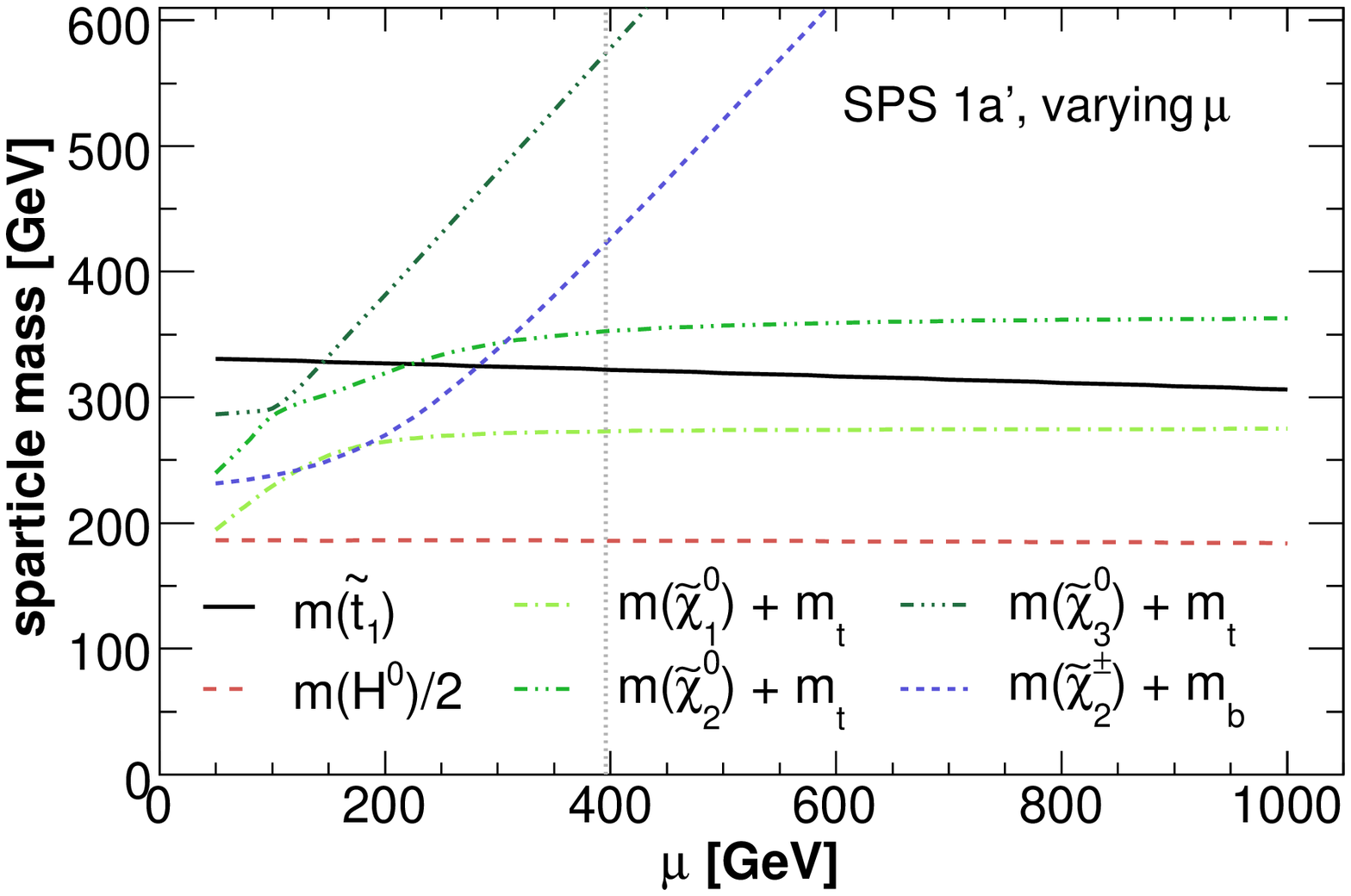}%
        \caption{Same as \figref{fig_MSQ3_dep}, but for variation of the 
        Higgs parameter $\mu$.}
        \label{fig_Mu_dep}
}
\clearpage

\section{Conclusions}

We have completed the NLO calculation for the $\stst$ production at hadron
colliders by providing the complete EW corrections at the one-loop level.

To obtain a consistent and IR-finite result, we have considered the 
interference terms between QCD and EW NLO terms for both virtual and
real contributions. Also, a new class of photon-induced
partonic processes of $\stst$ production occurs, 
which was found to yield considerable
contributions, comparable in size to the corrections to $\qq$ annihilation
and $gg$ fusion or even larger.

In total, the NLO EW contributions reach in size the 10-20\% level in the 
$\pt$ and invariant-mass
distributions and are thus significant.
Outside singular parameter configurations associated with thresholds,
the dependence on the MSSM parameters is rather smooth. 

\bigskip
Recently, a preprint appeared on the same topic~\cite{Beccaria:2007dt}, 
where the authors consider
virtual corrections and the soft part of the real corrections, 
both for the $gg$ fusion channel;  the hard part 
of the real corrections, as well as the
contributions from the other channels are missing. 
The numerical results can therefore not directly be compared with ours
at this stage.

\acknowledgments
The authors want to thank Tilman Plehn for helpful discussions
 and Edoardo Mirabella for cross-checking parts of the results.

\bigskip
\section*{Appendix}
\appendix

\addcontentsline{toc}{section}{Appendix}
\section{Feynman diagrams}

We show here generic Feynman Diagrams 
for the pair production of lighter top-squark 
at $\mathcal{O}(\alpha\alpha_s^2)$. 
Diagrams for $\tilde{t}_2^* \tilde{t}_2$ production can be constructed 
in complete analogy. The $q\bar{q}$ annihilation channels are exemplified 
by $u\bar{u}$ annihilation. Furthermore, the label $S^0$ refers to all 
neutral Higgs (and Goldstone) bosons $h^0,\,H^0,\,A^0,\,G^0$, 
and the label $S$ to all charged Higgs (and Goldstone) bosons $H^\pm,\, G^\pm$.
 
\vspace*{2cm}       

\FIGURE{
        \pict{.8}{0}{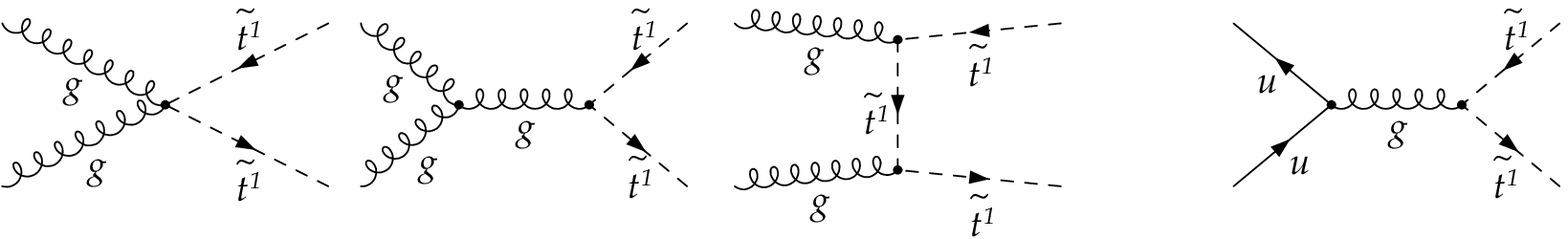}
        \caption{Feynman diagrams for top-squark pair production at 
         the Born level via 
        $gg$ fusion (left) and $\qq$ annihilation (right), here shown for $u$-quarks.  
        As in the following figures, diagrams with crossed final states are not shown explicitely.}
        \label{fig_born-beide}
}

\FIGURE{%
        \pict{.8}{0}{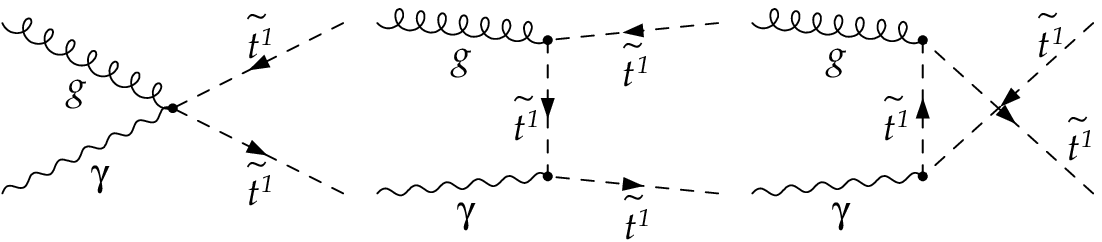}%
        \caption{Feynman diagrams for gluon-photon fusion.}
        \label{fig_gy}
}


\FIGURE{\small 
        \hspace*{2.5cm}\pict{.95}{0}{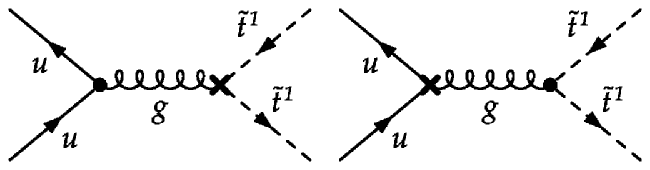}\\ \hspace*{3.5cm} (a) \\[1ex]
        \hspace*{.5cm}\pict{.95}{0}{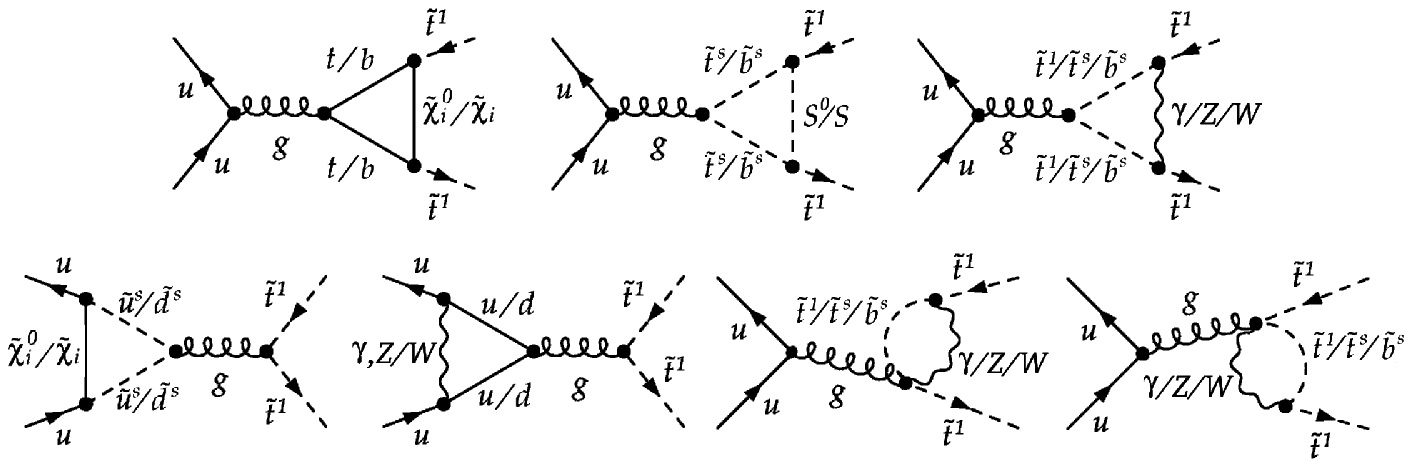}\\ \hspace*{3.5cm}  (b) \\[1ex]
        \hspace*{1cm}\pict{.95}{0}{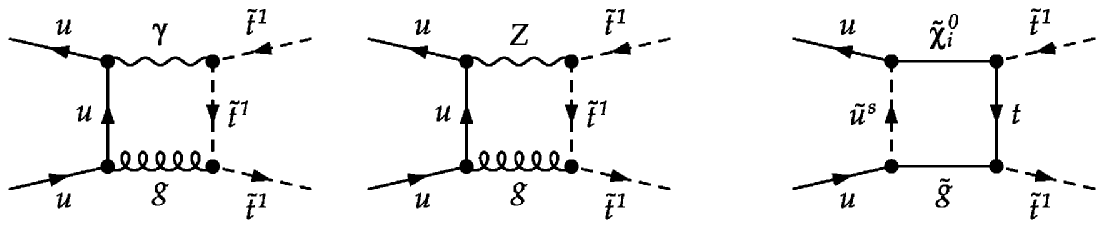}\\ \hspace*{1.5cm}(c) \hspace*{4.5cm} (d)
        \caption{Feynman diagrams for virtual corrections to 
        top-squark pair production via $q\bar{q}$ annihilation (here for $u$-quarks).
        The label $S^0$ refers to all neutral Higgs bosons $h^0,\,H^0,\,A^0,\,G^0$,
        the label $S$ to all charged Higgs bosons $H^\pm,\, G^\pm$.
        (a) counter-term diagrams, (b) vertex corrections, (c)~IR singular box diagrams,
        (d)~IR finite box diagram.}
        \label{fig_uu-virtualcorrs}
}


\FIGURE{\small
        \pict{.8}{0}{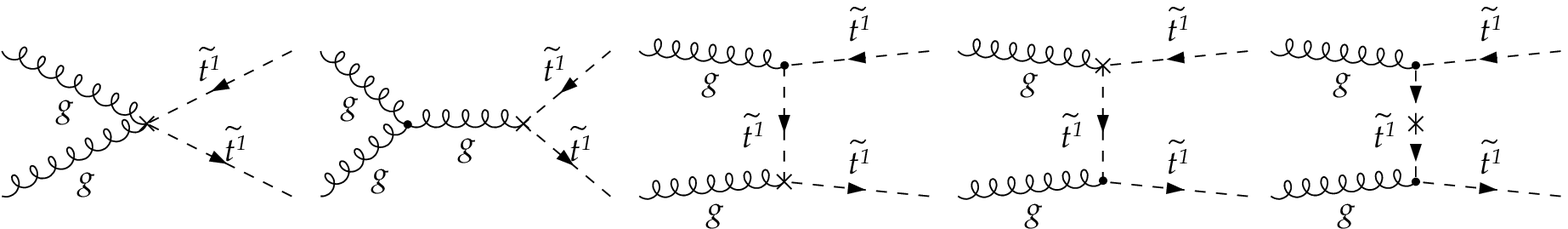}\\ (a)
        \pict{.8}{0}{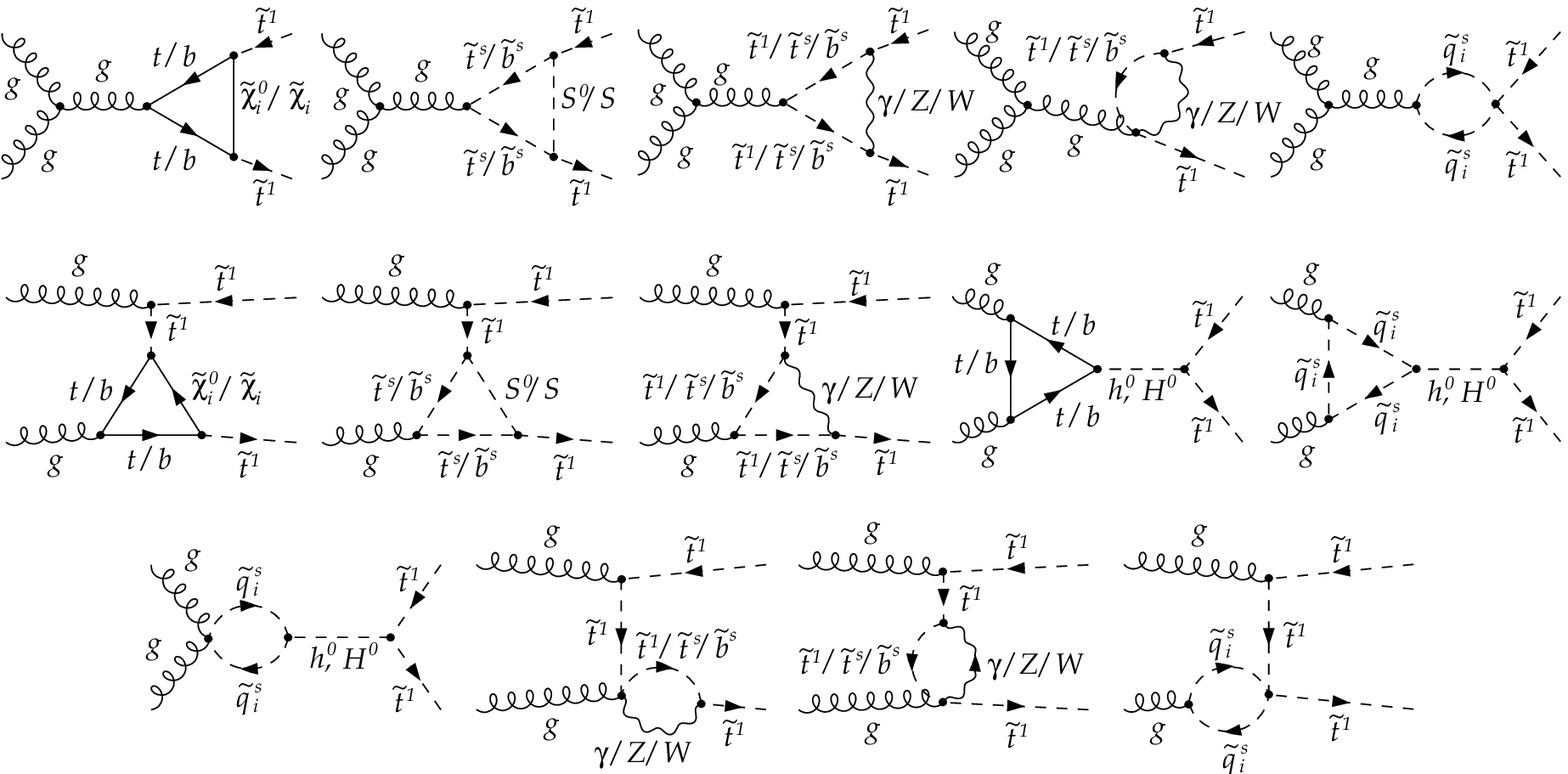}\\ (b)
        \pict{.8}{0}{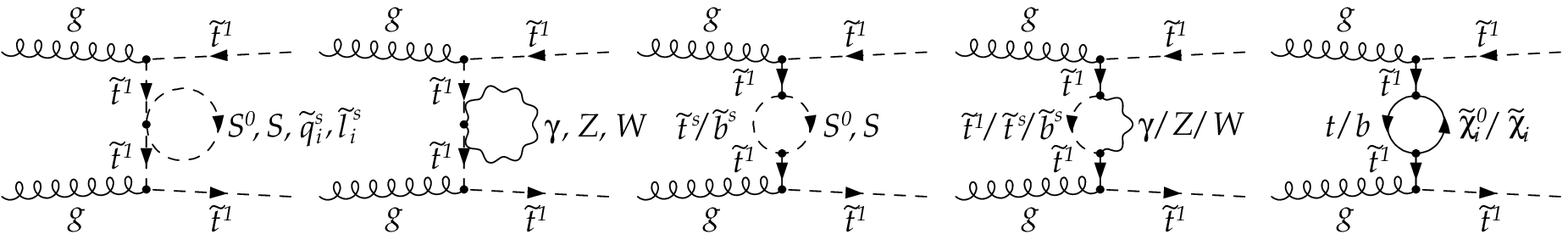}\ (c)
        \pict{.8}{0}{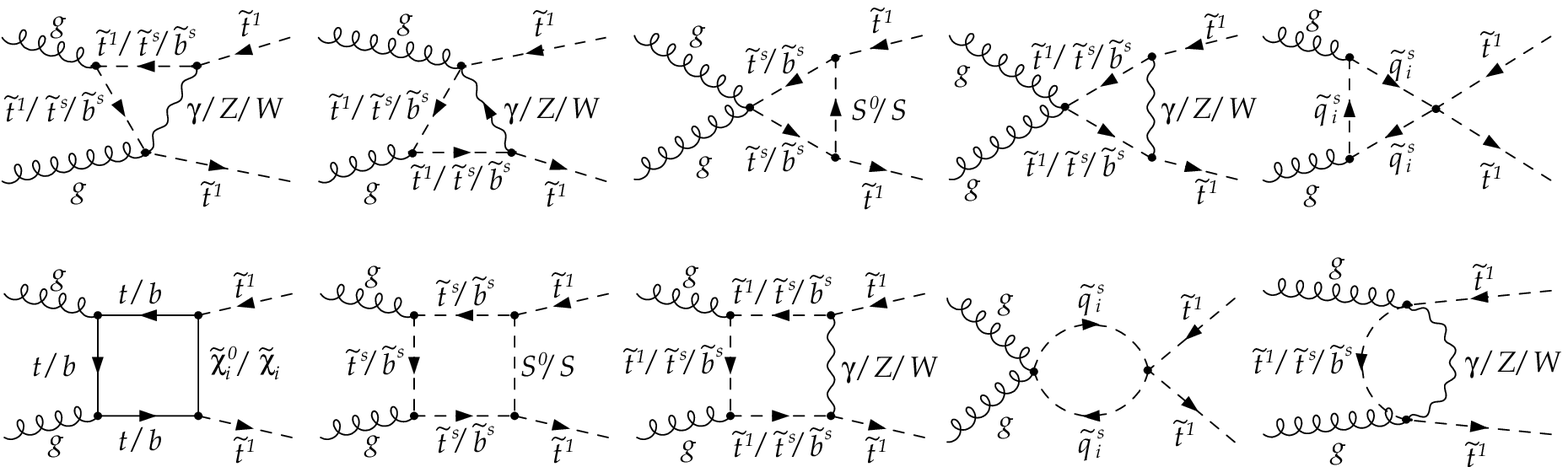}\\ (d)
        \caption{Feynman diagrams for virtual corrections to 
        top-squark pair production via $gg$ fusion, 
        diagrams with crossed final states are not explicitely shown.
        The label $S^0$ refers to all neutral Higgs bosons $h^0,\,H^0,\,A^0,\,G^0$,
        the label $S$ to all charged Higgs bosons $H^\pm,\, G^\pm$.
        (a) counter-term diagrams, (b) vertex corrections, (c) self-energy corrections, (d) box diagrams.%
        \label{fig_gg-virtualcorrs}}
}


\FIGURE{\small
        \hspace*{1.5cm}\pict{.8}{0}{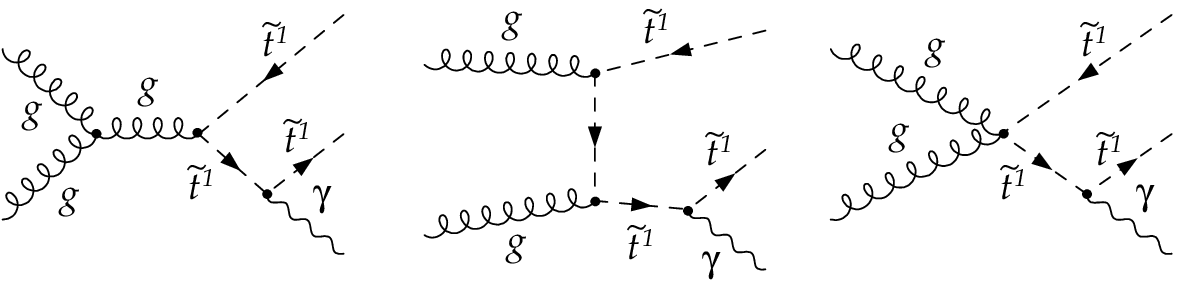} \\ \hspace*{3.2cm} (a) \\
        \hspace*{1.5cm}\pict{.8}{0}{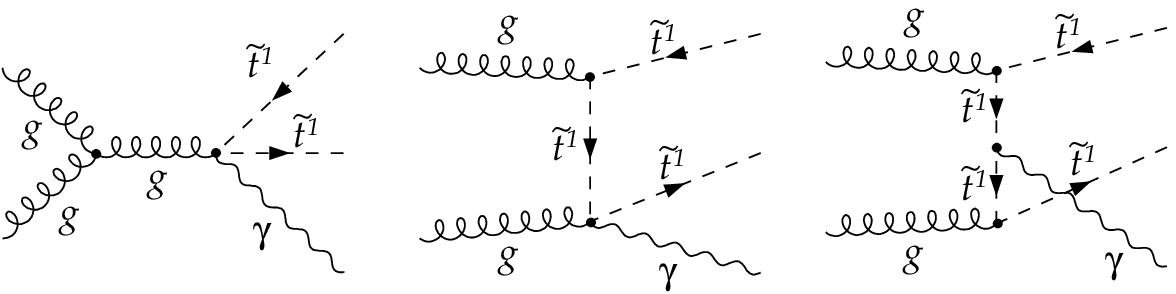}\\ \hspace*{3.2cm} (b) \\
        \hspace*{1.2cm}\pict{.8}{0}{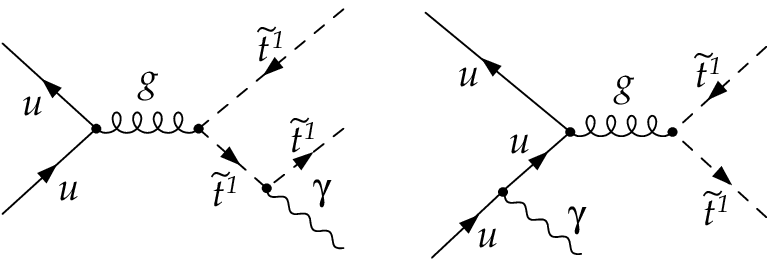} \quad \qquad
        \pict{.8}{0}{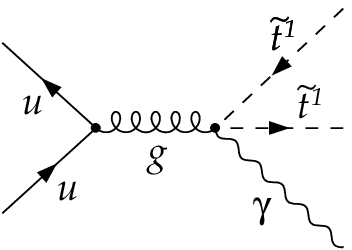} \\\hspace*{1.7cm} (c) \hspace*{5.cm} (d)
        \caption{Feynman diagrams for real photon radiation. (a) IR divergent -- (b) IR finite contributions for the $gg$ channel; (c) IR divergent -- (d) IR finite contributions for
        the $q\bar{q}$ channels. Feynman diagrams with photon radiation off the other quark or squark and with crossed final states are not shown explicitely.}
        \label{fig_photonrad}
}

\FIGURE{
   \pict{.8}{0}{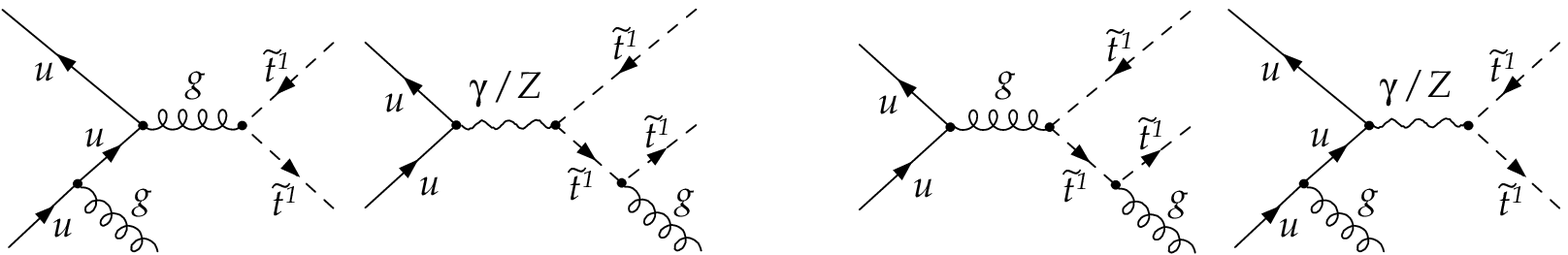}
        \caption{Feynman diagrams for gluon bremsstrahlung from the QCD and EW
                  Born diagrams (radiation from upper legs is not explicitly shown).
                  Only interference terms between initial and final state 
                  gluon radiation are non-vanishing.}
        \label{fig_gluonrad}
}

\FIGURE{%
        \pict{.8}{0}{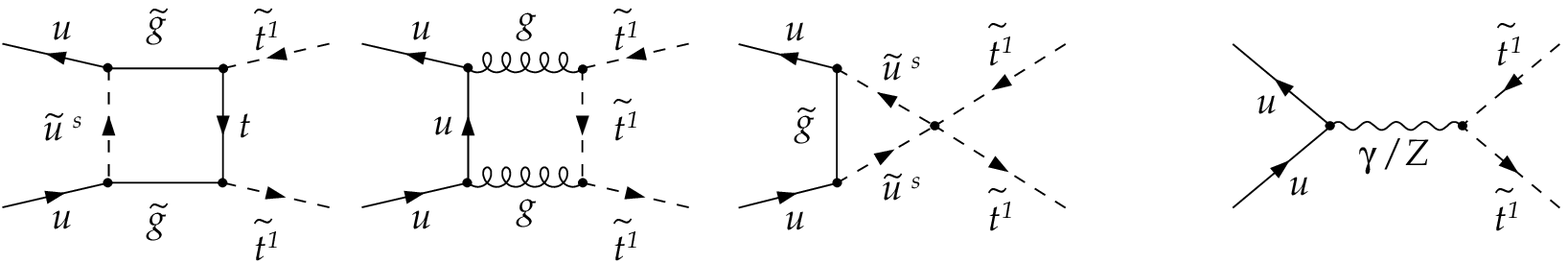}%
        \caption{Feynman diagrams for box contributions of 
		  $\mathcal{O}(\alpha_s^2)$ (left)
        interfering with electroweak Born graphs (right), here for
         $u\bar{u}$ annihilation.}
        \label{fig_uu-qcdBox}
}

\clearpage

\bibliographystyle{jhep}
\bibliography{draft} 
\end{document}